\documentclass[epj]{svjour}

\usepackage{graphics}
\usepackage{amssymb}
\usepackage{amsmath}
\usepackage{bm}
\usepackage{lineno}

\newcommand{\nn}{\nonumber}
\newcommand{\nslash}{\kern 0.2 em n\kern -0.50em /}
\newcommand{\kslash}{\kern 0.2 em k\kern -0.45em /}
\newcommand{\pslash}{\kern 0.2 em p\kern -0.50em /}
\newcommand{\Sslash}{\kern 0.2 em S\kern -0.50em /}
\newcommand{\Pslash}{\kern 0.2 em P\kern -0.50em /}
\newcommand{\Rslash}{\kern 0.2 em R\kern -0.50em /}
\newcommand{\Dslash}{\kern 0.2 em D\kern -0.65em /}

\begin{document}

\title{Di-hadron fragmentation and mapping of the nucleon structure}
\author{Silvia Pisano\inst{1} \and Marco Radici\inst{2}
}                     

\offprints{}          

\institute{INFN Laboratori Nazionali di Frascati, Via Enrico Fermi 40, 00044, Frascati, Italy \and INFN Sezione di Pavia, via Bassi 6, I-27100 Pavia, Italy}
\date{Received: date / Revised version: date}

\abstract{
The fragmentation of a colored parton directly into a pair of colorless hadrons is a non-perturbative mechanism that offers important insights into the nucleon structure. Di-hadron fragmentation functions can be extracted from semi-inclusive electron-positron annihilation data. They also appear in observables describing the semi-inclusive production of two hadrons in deep-inelastic scattering of leptons off nucleons or in hadron-hadron collisions. When a target nucleon is transversely polarized, a specific chiral-odd di-hadron fragmentation function can be used as the analyzer of the net density of transversely polarized quarks in a transversely polarized nucleon, the so-called transversity distribution. The latter can be extracted through suitable single-spin asymmetries in the framework of collinear factorization, thus in a much simpler framework with respect to the traditional one in single-hadron fragmentation. At subleading twist, the same chiral-odd di-hadron fragmentation function provides the cleanest access to the poorly known twist-3 parton distribution $e(x)$, which is intimately related to the mechanism of dynamical chiral symmetry breaking in QCD. When sensitive to details of transverse momentum dynamics of partons, the di-hadron fragmentation functions for a longitudinally polarized quark can be connected to the longitudinal jet handedness to explore possible effects due to $CP-$violation of the QCD vacuum. In this review, we outline the formalism of di-hadron fragmentation functions, we discuss different observables where they appear and we present measurements and future worldwide plans.
}  


\maketitle

\section{Introduction}
\label{intro}
In the hadronization process following an electron-positron annihilation, there is a non-vanishing probability that at the hard scale $Q^2$ of the process a highly virtual parton fragments directly into two hadrons inside the same jet with fractional energies $z_1$ and $z_2$, plus other unobserved fragments. This non-perturbative mechanism can be encoded in the so-called dihadron fragmentation functions (DiFFs) under the form $D(z_1, z_2; Q^2)$. The interest in two-particle correlations in $e^+ e^-$ processes was first pointed out in Ref.~\cite{Walsh:1974bj}, but DiFFs were introduced for the first time in the context of jet calculus~\cite{Konishi:1978yx}. DiFFs are also needed to cancel all collinear singularities when the semi-inclusive production of two back-to-back hadrons from $e^+ e^-$ annihilations is considered at next-to-leading order (NLO) in the strong coupling constant~\cite{deFlorian:2003cg}. 

Experimental information on two hadron production is often delivered in terms of a distribution in the invariant mass $M_h$ of the hadron pair~\cite{Acton:1992sa,Abreu:1992xx,Buskulic:1995gm}. Therefore, it is convenient to describe the process with ``extended" DiFFs of the form 
$D(z_1, z_2, M_h; Q^2)$, in analogy to what is done for fracture functions~\cite{Grazzini:1997ih}. Recently, the problem of two-hadron production when one hadron is in the current fragmentation region and one in the target region has also been considered~\cite{Anselmino:2011bb}. If $M_h^2 \approx Q^2$, DiFFs transform into the convolution of two single-hadron fragmentation functions~\cite{Zhou:2011ba}. If $M_h^2 \ll Q^2$, they represent a truly new  non-perturbative object. The definition of DiFFs and a thorough study of their properties were presented in Refs.~\cite{Bianconi:1999cd,Bacchetta:2002ux} (up to leading twist) and in Ref.~\cite{Bacchetta:2003vn} (including subleading twist). At $M_h^2\ll Q^2$, DiFFs satisfy the same evolution equations as the single-hadron fragmentation functions in collinear kinematics~\cite{Ceccopieri:2007ip}, in contrast to what happens if DiFFs are integrated over $M_h^2$~\cite{deFlorian:2003cg}. They can be factorized and are assumed to be universal. In fact, they appear not only in $e^+ e^-$ annihilations~\cite{Boer:2003ya,Bacchetta:2008wb,Courtoy:2012ry}, but also in hadron pair production in semi-inclusive deep-inelastic scattering (SIDIS)~\cite{Bacchetta:2002ux,Bacchetta:2008wb} and in hadronic collisions~\cite{Bacchetta:2004it}. 

For polarized fragmentations, certain DiFFs emerge from the interference of amplitudes with the hadron pair being in two states with different relative angular 
momentum~\cite{Collins:1994ax,Jaffe:1998hf,Radici:2001na,Bacchetta:2006un}. Hence, in the literature they are addressed also as interference fragmentation functions (IFFs)~\cite{Jaffe:1998hf}. IFFs can be used in particular as analyzers of the polarization state of the fragmenting parton~\cite{Bianconi:1999cd,Efremov:1992pe,Collins:1994kq,Artru:1995zu}. In SIDIS on transversely polarized targets, IFFs have become popular because they allow to extract in a simple framework the so-called transversity parton distribution function, which describes the balance between number densities of partons with transverse polarization aligned or antialigned to the transverse polarization of the parent nucleon (for a review on transversity, see Ref.~\cite{Barone:2001sp} and references therein). 

At leading twist, the transversity is the only one, out of four chiral-odd TMDs, that survives after integrating upon the parton transverse momenta. Hence, it shares the same relevance as the momentum and helicity distributions, and together with them it gives a complete description at leading twist of the spin structure of spin-half hadrons in collinear kinematics. Its first moment gives the tensor charge, which can represent a useful testing ground for searches of new physics beyond the Standard Model (see Ref.~\cite{Courtoy:2015haa} and references therein). Being chiral-odd, transversity can be measured only in processes with two hadrons in the initial state, or one hadron in the initial state and at least one hadron in the final state ({\it e.g.} SIDIS). Transversity was extracted for the first time by combining data on polarized single-hadron SIDIS together with data on almost back-to-back emission of two hadrons in $e^+ e^-$ annihilations~\cite{Anselmino:2008jk,Anselmino:2013vqa}. The difficult part of this analysis lies in the factorization framework used to interpret the data, since it involves TMDs. QCD evolution of TMDs must be included to analyze SIDIS and $e^+ e^-$ data obtained at very different scales, and an attempt to give a complete description of these effects was only recently released~\cite{Kang:2014zza,Kang:2015msa}.   

As anticipated above, the DiFFs offer a simpler alternative route to transversity by using SIDIS with two hadrons detected in the final state in the standard framework of collinear factorization, namely when there is no sensitivity to the transverse dynamics of partons. In this case, at leading twist the cross section contains a contribution where the chiral-odd transversity is multiplied to a specific chiral-odd IFF named $H_1^{\sphericalangle}$~\cite{Jaffe:1998hf,Radici:2001na,Collins:1994kq}, which in turn can be extracted from the corresponding $e^+ e^-$ annihilation process leading to two back-to-back hadron pairs~\cite{Boer:2003ya,Courtoy:2012ry}. At subleading twist, the cross section displays other structures where the (polarized) DiFFs can be useful analyzers of interesting functions. For example, from beam-spin asymmetries it is possible to isolate a term involving the same $H_1^{\sphericalangle}$ and the twist-3 chiral-odd distribution $e(x)$~\cite{Bacchetta:2003vn}, related to the mechanism of the spontaneous breaking of QCD chiral symmetry and, ultimately, to the strange-quark content of the nucleon~\cite{Jaffe:1991kp}.

From this short introduction, it emerges that DiFFs are convenient tools to access  elusive/suppressed parton distribution functions that are necessary, however, to improve our mapping of the spin structure of the nucleon. Furthermore, this access is granted in a simple framework (at least, from the theoretical point of view) where the kinematics is collinear, namely with no manifest dependence on the parton transverse momenta. Anyway, it is useful to include such dependence and explore the whole formalims of the TMD DiFFs~\cite{Bianconi:1999cd,Gliske:2014wba}. In fact, in the cross section we can either come across terms that are similar to the single-hadron fragmentation case (and whose measurement can represent an important cross-check of the elementary mechanism described by the corresponding TMD PDF), or we can find new contributions that have no such counterpart and that represent, therefore, a new window on the non-perturbative phenomena happening during fragmentation. 

In the following, we describe the general formalism about DiFFs in Sec.~\ref{sec::theory}. In Sec.~\ref{sec::h1_extr}, we describe the extraction of DiFFs from $e^+ e^-$ annihilation data and recall the main steps for extracting the transversity from two-hadron SIDIS data and proton-proton collision data in the framework of collinear factorization, giving also some perspectives about future developments and measurements. In Sec.~\ref{sec::twist3}, we extend the DiFF formalism at subleading twist, discussing the possible access in the collinear framework to the interesting PDF $e(x)$ through present and future measurements of the related spin asymmetry. In Sec.~\ref{sec::TMDDiFF}, we describe some interesting applications when DiFFs are considered also as functions of parton transverse momenta, like the possibility of connecting the helicity DiFF to the longitudinal jet handedness function. Finally, in Sec.~\ref{sec::conclusions} we summarize and discuss some outlooks. 

\begin{figure}[h]
\begin{center}
\resizebox{0.25\textwidth}{!}{%
  \includegraphics{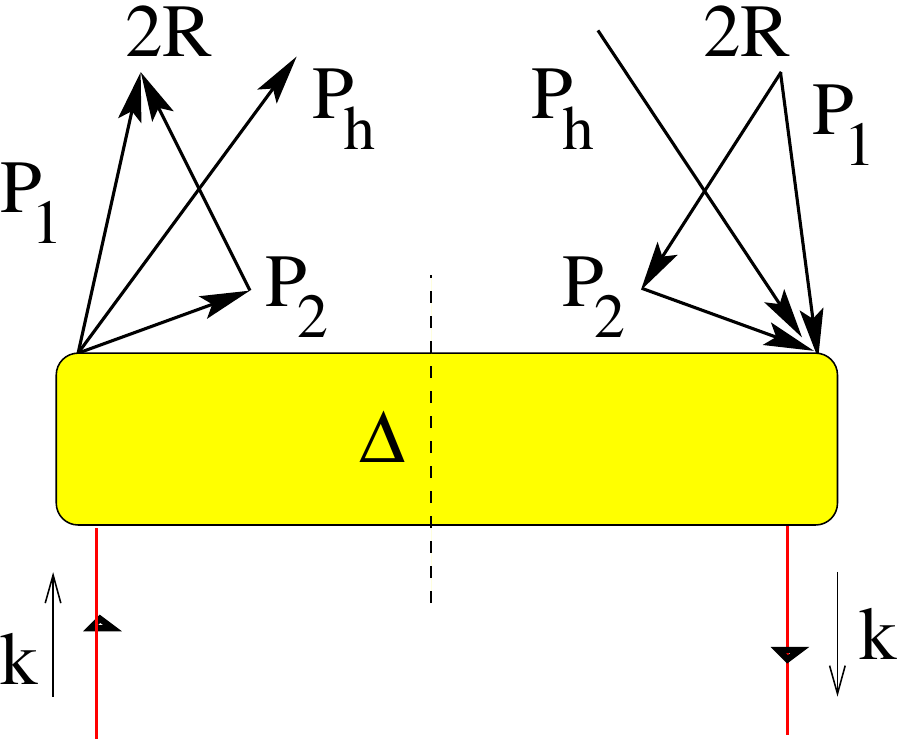}
}
\caption{Quark-quark correlation function $\Delta$ for the fragmentation of a quark with momentum $k$ into a pair of hadrons with total momentum $P_h = P_1 + P_2$ and relative momentum $R = (P_1 - P_2) / 2$.}
\label{fig::2hcorrelator}       
\end{center}
\end{figure}

\section{General Formalism}
\label{sec::theory}

The fragmentation process is schematically depicted in Fig.~\ref{fig::2hcorrelator}: a quark with momentum $k$ and mass $m$ fragments into two unpolarized hadrons with momenta $P_1, P_2$ and masses $M_1, M_2$. We introduce the pair total momentum $P_h = P_1+P_2$ and the pair relative momentum $R = (P_1 - P_2)/2$. It is convenient to describe the fragmentation in the frame where $\bm{P}_{hT}=0$. We define the following kinematic invariants
\begin{equation}
\centering
z = \frac{P_h^-}{k^-}\equiv z_1 + z_2 \; ,
\quad
\zeta = \frac{2 R^-}{P_h^-} = \frac{z_1 - z_2}{z}  \; ,
 \label{e:zetaz}
\end{equation}
where $z_1, \, z_2,$ are the fractional energies carried by the two final hadrons. The light-cone components of a 4-vector are obtained by projecting it along two light-like directions $n_+$ and $n_-$ satisfying $n_+^2 = n_-^2 = 0$ and $n_+ \cdot n_- = 1$. 

The quark-quark correlator of Fig.~\ref{fig::2hcorrelator} can be written at leading twist  as~\cite{Bacchetta:2003vn}:
\begin{equation}
\begin{split}
\Delta (z, &\zeta, \bm{R}_T^2, \bm{k}_T^2, \bm{k}_T\cdot \bm{R}_T) = \\
&\frac{1}{16 \pi}\, \bigg\{ D_1 \,\nslash_- +
H_1^{\sphericalangle}\, \frac{\mathrm{i}}{M_h}\,\Rslash^{}_T\, \nslash_- \\ 
&\  + H_1^{\perp}\, \frac{\mathrm{i}}{M_h}\,\kslash^{}_T\, \nslash_- + 
G_1^{\perp} \, \frac{\epsilon_T^{\rho \sigma} R^{}_{T \rho} k^{}_{T \sigma}}{M_h^2}\, \gamma_5 \nslash_-\bigg\} \; ,  
\end{split} 
\label{eq::decomDelta}
\end{equation}
where $\epsilon_T^{\mu \nu} = \epsilon^{\rho \sigma \mu \nu} n_{ + \rho}\, n_{- \sigma}$. The DiFFs $D_1, G_1^\perp, H_1^{\sphericalangle}, H_1^\perp,$ are all functions of $z, \zeta, \bm{R}_T^2, \bm{k}_T^2, \bm{k}_T\cdot \bm{R}_T$. They can be projected out of the correlator $\Delta$ by suitable Dirac structures that correspond to specific polarization states of the fragmenting quark. Correspondingly, the DiFFs have nice probabilistic interpretations~\cite{Bianconi:1999cd}: they all concern two unpolarized hadrons emerging from the same fragmentation, but $D_1$ describes the process as initiated from an unpolarized quark, $G_1^\perp$ describes the balance between densities for  longitudinally polarized initiating quarks with opposite helicities, $H_1^{\sphericalangle}$ and $H_1^\perp$ describe the same balance but for transversely polarized quarks. $G_1^\perp, H_1^{\sphericalangle}, H_1^\perp$ are (na\"ive) $T-$odd; $D_1$ and $G_1^\perp$ are chiral-even, while $H_1^{\sphericalangle}$ and $H_1^\perp$ are chiral-odd. The $H_1^\perp$ is the analogue of the Collins function for the single-hadron fragmentation case.

If we restrict to collinear kinematics and we integrate over the $\bm{k}_T$ dependence, only the $D_1$ and $H_1^{\sphericalangle}$ terms survive in Eq.~(\ref{eq::decomDelta}) and become functions of $z, \zeta, \bm{R}_T^2$. Then, the  probability density for finding a pair of unpolarized hadrons $(h_1, h_2)$ generated by a transversely polarized quark $q^\uparrow$ becomes
\begin{equation}
D_{(h_1,h_2)/q^\uparrow} (z, \zeta, \bm{R}_T^2, \phi_R) = D_1^q - H_1^{\sphericalangle\, q} \frac{\bm{S}_q \cdot (\hat{\bm{k}} \times \bm{R}_T)}{M_h} \; , 
\label{eq::H1angle_dens}
\end{equation}
where $\hat{\bm{k}}$ represents the direction of the fragmenting quark momentum and we have $\hat{\bm{k}} \equiv \bm{P}_h$. 

\begin{figure}[h]
\begin{center}
\resizebox{0.45\textwidth}{!}{%
  \includegraphics{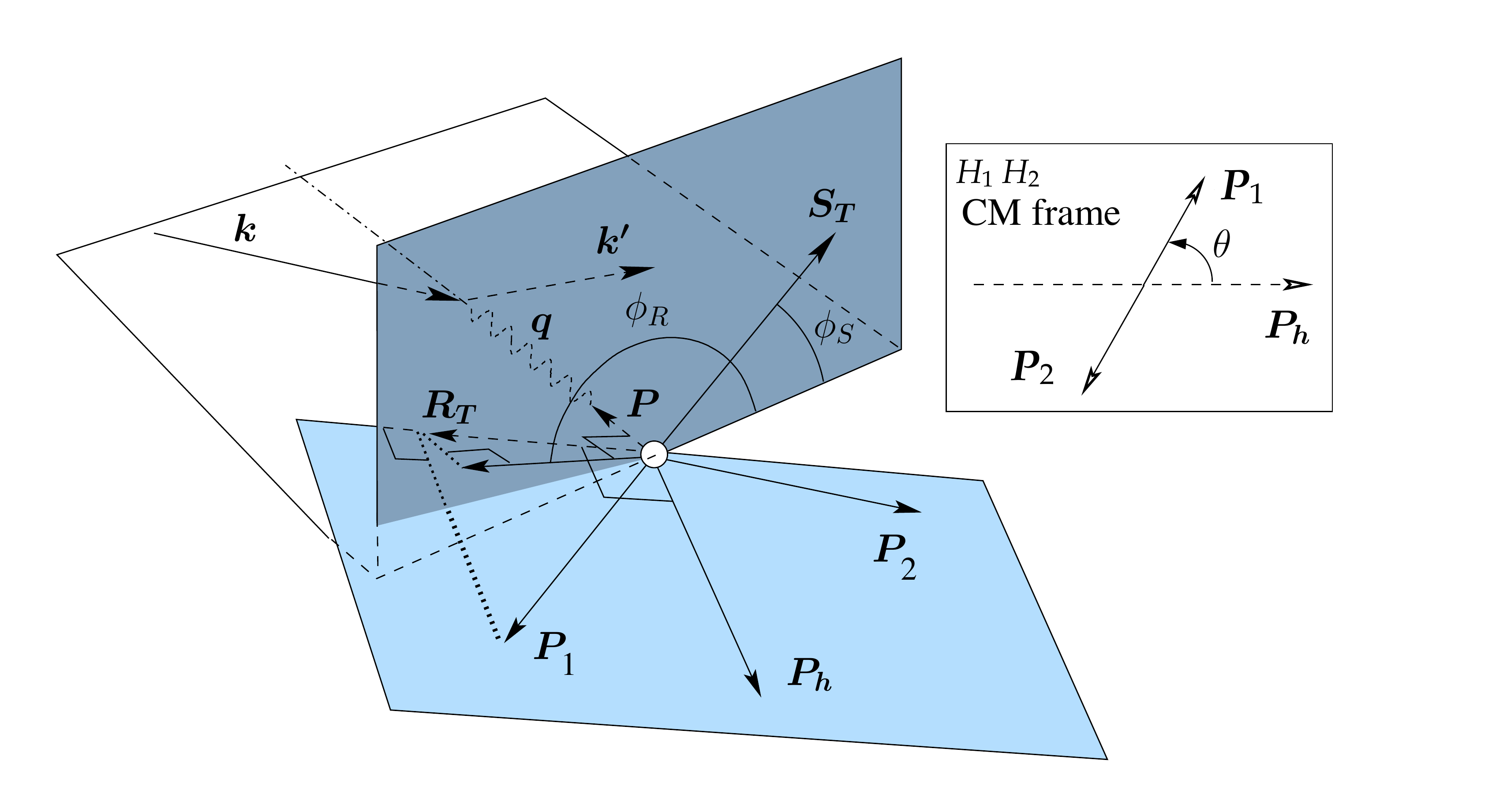}
}
\caption{Kinematics of the two-hadron semi-inclusive production in lepton scattering off a polarized target. The azimuthal angle $\phi_{R}$ refers to the component of $\bm{R}_T$ which is transverse to the virtual-photon-target $(\bm{q}, \bm{P})$ plane. Similarly for $\phi_S$ referred to the transverse polarization vector $\bm{S}_T$.}
\label{fig::SIDISkin}       
\end{center}
\end{figure}

When the pair invariant mass $P_h^2 = M_h^2$ is small compared to the hard scale of the process, the hadron pair can be assumed to be produced mainly in relative $s$ or $p$ waves, suggesting that the DiFFs can be conveniently expanded in partial waves. In the center-of-mass (cm) frame of the two hadrons, the emission occurs back-to-back and the key variable is the angle $\theta$ between the direction of emission and $P_h$ (see Fig.~\ref{fig::SIDISkin}). It turns out that~\cite{Bacchetta:2002ux}
\begin{align}
\bm{R}_T &= \bm{R} \sin \theta \; ,  \nn \\
\vert \bm{R} \vert &= \frac{1}{2}\, \sqrt{M_h^2 - 2(M_1^2+M_2^2) + (M_1^2-M_2^2)^2/M_h^2} 
\; , 
\label{eq::Rvect}
\end{align}
and that $\zeta$ can be shown to be a linear polynomial in $\cos \theta$. Then, DiFFs can be expanded in Legendre polynomials in $\cos \theta$~\cite{Bacchetta:2002ux}:
\begin{align}
D_1 &\rightarrow  D_{1, ss+pp} + D_{1, sp} \cos \theta + D_{1, pp} \frac{1}{4} (3\cos^2\theta -1) \; ,  \nn \\
\frac{|\bm{R}_T|}{M_h}\, H_1^{\sphericalangle} &\rightarrow H_{1, sp}^{\sphericalangle}\, \sin\theta + H_{1, pp}^{\sphericalangle} \, \sin\theta\, \cos\theta \; , 
\label{eq::LMexp}
\end{align}
where each term with a specific partial wave is function of $z, M_h^2$. After averaging over $\cos \theta$, only the terms $D_{1, ss+pp}$ and $H_{1, sp}^{\sphericalangle}$ survive in the expansion. The former corresponds to an unpolarized quark fragmenting into an unpolarized pair being created in a relative $\Delta L=0$ state. The latter relates the transverse polarization of the fragmenting quark to the interference of unpolarized hadron pairs produced with $|\Delta L| = 1$. The simplification holds even if the $\theta$ dependence in the acceptance is not complete but symmetric about $\theta = \pi / 2$. Without ambiguity, the two surviving terms will be identified with $D_1$ and $H_1^{\sphericalangle}$, respectively. A similar partial-wave expansion holds also for DiFFs at subleading twist~\cite{Bacchetta:2003vn}.
%
%
%
%
\section{DiFFs and the extraction of transversity}
\label{sec::h1_extr}
The non-perturbative mechanism described by the correlation in Eq.~(\ref{eq::H1angle_dens}) represents the alternative to the Collins effect for extracting the transversity distribution. It relates the transverse polarization of the fragmenting parton with flavor $q$ to the azimuthal orientation of the plane containing the momenta of the detected hadron pair, identified by the azimuthal angle $\phi_{R_T}$ of the vector $\bm{R}_T$. 
%
\subsection{The target-spin asymmetry in SIDIS}
\label{sec::SIDISssa}
In fact, at leading order in the couplings the differential cross section for the two-hadron SIDIS of an unpolarized lepton with momentum $\ell$ off a nucleon target with momentum $P$ and transverse polarization $|\bm{S}_T|$ reads~\cite{Radici:2001na,Bacchetta:2012ty,Radici:2015mwa}
\begin{align}
\lefteqn{\frac{d\sigma}{dx \, dy\, \,dz\, d\phi_{R}\, d M_h^2} = 
\frac{\alpha^2}{x y\, Q^2} }  \nn \\ 
&\  \times \Bigg\{ A(y) \, 2 \, \sum_q e_q^2\, f_1^q(x; Q^2)\, D_1^q \left(z, M_h^2; Q^2 \right)  \nn \\
&\quad - |\bm{S}_T|\, B(y) \, \sin(\phi_{R}+\phi_{S})\, \frac{\pi}{2} \, \frac{|\bm{R}|}{M_h} \nn \\
&\qquad  \times \sum_q e_q^2\,  h_1^q(x; Q^2)\,H_1^{\sphericalangle\, q}\left(z, M_h^2; Q^2 \right) \Bigg\} \; ,
\label{eq::SIDIScross}
\end{align}
where $\alpha$ is the fine structure constant, $A(y) = 1 - y + y^2/2$, $B(y) = 1 - y$, 
$\phi_S = \pi / 2$, $e_q$ is the fractional charge of a parton with flavor $q$, $Q^2$ is the hard scale of the SIDIS process with spacelike momentum transfer $q^2 = -Q^2$, and the usual invariants are defined as $x= Q^2/ 2P\cdot q$ and $y = P\cdot q / P\cdot \ell$.  

The angle $\phi_R$ in Eq.~(\ref{eq::SIDIScross}) is not the same as the angle $\phi_{R_T}$ that describes the azimuthal orientation of the plane with the di-hadron momenta in Eq.~(\ref{eq::H1angle_dens}). In a SIDIS process, two different transverse projections can be considered: with respect to the $(P,P_h)$ plane or to the $(P,q)$ plane (see Fig.~\ref{fig::SIDISkin}). The vector $R_T$ described in Sec.~\ref{sec::theory} is the transverse component of $R$ with respect to the $(P, P_h)$ plane, and $\phi_{R_T}$ is the azimuthal angle of its spatial component $\bm{R}_T$. However, the cross section depends on the azimuthal angle of $R_T$ with respect to the $(P, q)$ plane that we indicate with $\phi_R$. In Ref.~\cite{Gliske:2014wba}, a covariant definition of $\phi_R$ is given starting from the covariant definition of $R_T$. It is shown that this definition coincides up to $1/Q^2$ corrections with all the non-covariant definitions adopted in the literature before, in particular for the experimental measurements described here below. For convenience, the explicit expression of $\phi_R$ in the target rest frame (or in any frame reached from the target rest frame by a boost along $\bm{q}$) is 
\begin{equation}
\phi_{R} =  
\frac{(\bm{q} \times \bm{\ell}) \cdot \bm{R}_T}{\vert (\bm{q} \times \bm{\ell}) \cdot \bm{R}_T\vert}  
\arccos {
\frac{(\bm{q} \times \bm{\ell}) \cdot (\bm{q}\times \bm{R}_T)}
        {\vert \bm{q} \times \bm{\ell}\vert \   \vert \bm{q}\times \bm{R}_T\vert }   } \; . 
\label{eq::phiR}
\end{equation}

From Eq.~(\ref{eq::SIDIScross}), we can define the following target-spin asymmetry~\cite{Radici:2001na,Bacchetta:2012ty,Radici:2015mwa}:
\begin{align}
\lefteqn{A_{\mathrm{SIDIS}} (x, z, M_h; Q) =} \nn \\ 
&\quad  \frac{1}{|\bm{S}_T|}\, \frac{\frac{8}{\pi} \, \int d\phi_R \, \sin (\phi_R+\phi_S)\, (d\sigma^\uparrow - d\sigma^\downarrow)}{\int d\phi_R \, (d\sigma^\uparrow + d\sigma^\downarrow)} \nn \\
&= - \frac{B(y)}{A(y)} \,\frac{|\bm{R}|}{M_h} \, 
\frac{\sum_q\, e_q^2\, h_1^q (x; Q^2)\, H_1^{\sphericalangle\, q}(z, M_h^2; Q^2)} 
        {\sum_q\, e_q^2\, f_1^q(x; Q^2) \, D_{1}^q(z, M_h^2; Q^2)} \; ,
\label{eq::SIDISssa}
\end{align} 
which is proportional to the product of the transversity $h_1$ and the IFF $H_1^{\sphericalangle}$, and not to a convolution on parton transverse momenta, as it happens in the Collins effect. This is a direct consequence of the fact that the correlation 
$\bm{S}_q \cdot (\hat{\bm{k}} \times \bm{R}_T)$ in Eq.~(\ref{eq::H1angle_dens}) produces an asymmetric azimuthal modulation in the cross section also in collinear kinematics. No assumptions are necessary about the dependence of $h_1$ and $H_1^{\sphericalangle}$ on the transverse momenta of partons. As such, the measurement of $A_{\mathrm{SIDIS}}$ provides a model-independent cross-check to the extraction of transversity from the Collins effect in single-hadron fragmentation, provided that the unknown DiFFs are independently extracted from another process.

\subsection{The HERMES measurement}
\label{sec::HERMES}

The first observation of a non-zero $A_{\mathrm{SIDIS}}$ was reported by the \texttt{HERMES} collaboration~\cite{hermes_2008}. The analysis was performed on a data set collected by impinging a $e^\pm$ beam of 27.6 GeV on a gaseous hydrogen target transversely polarized (with an average target polarization $\left\langle S_T\right\rangle$=0.74). The final sample of $\pi^+ \pi^-$ was selected by removing the resonance region through the cut $W^2 > 10$ GeV$^2$, with $W$ the invariant mass of the virtual-photon-nucleon system. The deep-inelastic regime was selected by requiring $Q^2 > 1$ GeV$^2$, and the cut $0.1 < y < 0.85$ removed the kinematics where radiative effects could be dominant, that lies in the high-$y$ region. The contributions from exclusive two-pion electro-production were excluded by requiring a missing mass $M_X > 2$ GeV. In order to select pions coming from the struck quark fragmentation, a minimum momentum cut $P_{\pi}>1$ GeV was applied to identify final hadrons. 

\begin{figure}[h]
\begin{center}
\resizebox{0.45\textwidth}{!}{%
  \includegraphics{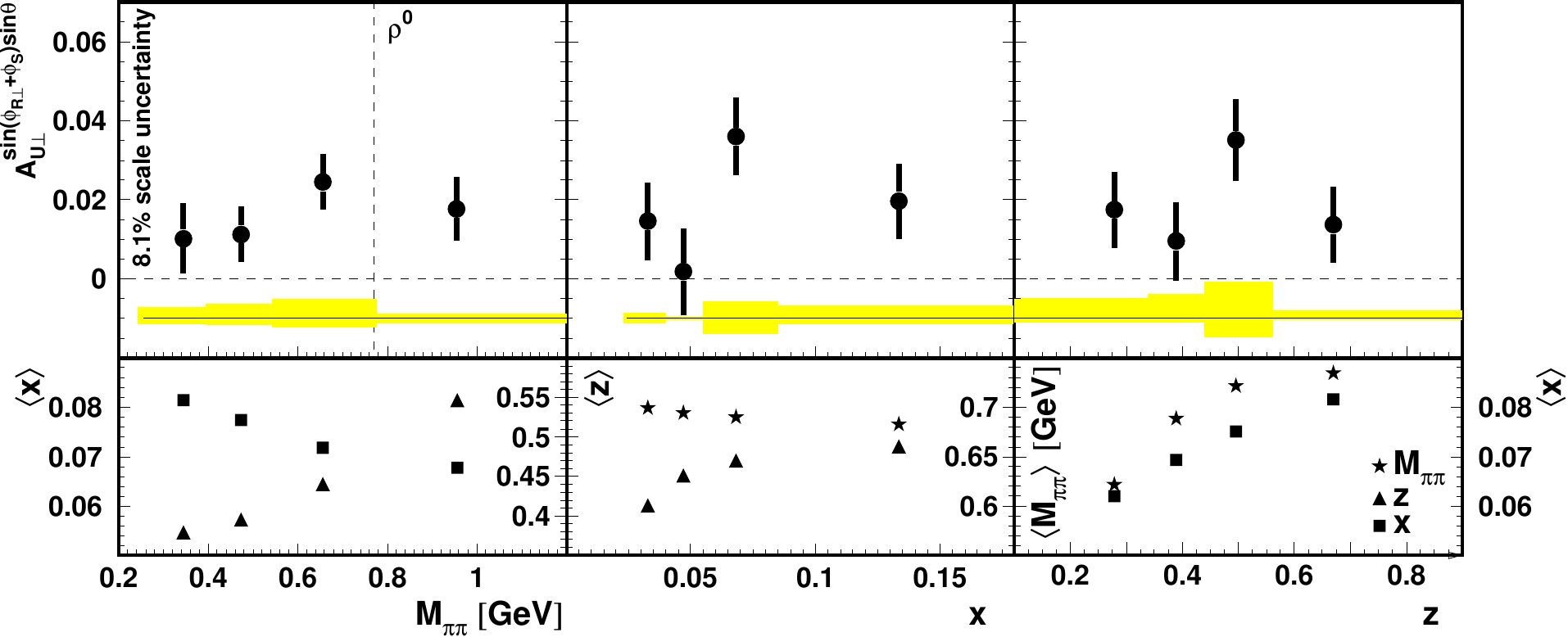}
}
\caption{The target-spin asymmetry for semi-inclusive $\pi^+ \pi^-$ production off a transversely polarized proton measured by the \texttt{HERMES} Collaboration as a function of the pair invariant mass $M_{\pi \pi}$, of $x$ and $z$~\cite{hermes_2008}. The bottom panel shows the average value of the integrated variables. In the $x$ and $z$ dependence $M_{\pi\pi}$ is limited to the range 0.5$\div$1.0 GeV. A scale uncertainty of 8.1\% has to be added to account for the uncertainty in the target polarization. Other systematic uncertainties are summed in quadrature and displayed by the asymmetric error band.}
\label{fig::hermes_results}       
\end{center}
\end{figure}


\vspace{-0.5cm}
Experimentally, $A_{\mathrm{SIDIS}}$ is defined as
\begin{equation}
A_{\mathrm{SIDIS}} (x, z, M_{\pi\pi}) \equiv \frac{1}{|\bm{S}_T|} \, 
\frac{N^{\uparrow} -  N^{\downarrow}}{N^{\uparrow} + N^{\downarrow}} \; , 
\label{eq:SIDISexpssa}
\end{equation}
where $N^{\uparrow (\downarrow)}$ refers to the number of events collected for a target polarization with $\phi_S = \pi / 2 \  (\phi_S = 3 \pi / 2)$ (in reality, the target spin direction is relative to the incoming lepton beam, but in deep-inelastic kinematics the latter can be safely  replaced with the virtual-photon direction~\cite{Diehl:2005pc}). The asymmetry is measured as a function of $x, z, M_{\pi\pi}\equiv M_h$, and summed over $\phi_R$ and $\theta$. The results are shown in Fig.~\ref{fig::hermes_results}. They corresponds to an average kinematics of $\left\langle x\right\rangle$ = 0.07, $\left\langle y\right\rangle$ = 0.64, $\left\langle Q^2\right\rangle$ = 2.35 GeV$^2$, $\left\langle z\right\rangle$ = 0.43. 

According to the Trento conventions~\cite{Bacchetta:2004jz}, the asymmetry turned out to be positive over the whole range:  the transversity and IFF are different from zero in the explored kinematics, and from Eq.~(\ref{eq::SIDISssa}) we deduce that most likely they have opposite sign flavor by flavor. Since, {\it e.g.}, the transversity for the up quark is known from the Collins effect to be positive, a negative IFF $H_1^{\sphericalangle\, u}$ in Eq.~(\ref{eq::H1angle_dens}) indicates that an up quark moving along the $\hat{\bm{z}}$ axis and polarized along $\hat{\bm{y}}$ fragments in a pair with a $\pi^+$ preferentially emitted along $\hat{\bm{x}}$ and a $\pi^-$ along $-\hat{\bm{x}}$ (if $\bm{R}$ conventionally points to the positively charged particle in the pair). 

\subsection{The COMPASS measurement}
\label{sec::COMPASS}

\begin{figure}[h]
\begin{center}
\resizebox{0.45\textwidth}{!}{%
  \includegraphics{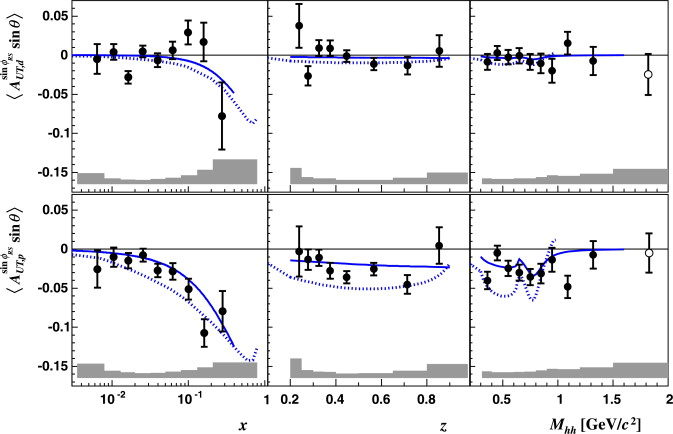}
}
\caption{The target-spin asymmetry for semi-inclusive unidentified $h^+ h^-$ production off a transversely polarized target measured by the \texttt{COMPASS} Collaboration as a function of $x$, $z$, and of the pair invariant mass $M_{hh}\equiv M_h$~\cite{compass_2012}. Upper panel for the deuteron target, lower panel for the proton. The grey bands indicate the systematic uncertainties. Solid lines show the predictions based on the spectator model of Ref.~\cite{Bacchetta:2006un} and on the transversity extracted from the Collins effect in Ref.~\cite{Anselmino:2008jk}, dotted lines refer to an analysis based on the pQCD counting rule~\cite{she_2008}.}
\label{fig::compass_results_2012}
\end{center}
\end{figure}

A second measurement of $A_{\mathrm{SIDIS}}$ was performed by the \texttt{COMPASS} collaboration~\cite{compass_2012}. Data were collected by letting the 160-GeV muon beam produced at the CERN SPS hit solid NH$_3$ and $^6$LiD targets with average transverse polarization $\left\langle S_T\right\rangle = 0.83$ and $\left\langle S_T\right\rangle = 0.47$, respectively. SIDIS events with the production of unidentified $h^+h^-$ pairs were selected through the cuts $W^2 > 25$ GeV$^2$, $Q^2 > 1$ GeV$^2$, $0.1 < y < 0.9$ and $M_X > 2.4$ GeV. Hadrons coming from the actual fragmentation of the struck quark are selected through the cuts $z > 0.1$ and $x_F > 0.1$.  The pair invariant mass was limited to $M_h < 1.5$ GeV in order to justify the inclusion of only relative $s$ and $p$ waves in the DiFF partial-wave expansion, as in Eq.~(\ref{eq::LMexp}). In Fig.~\ref{fig::compass_results_2012}, the target-spin asymmetry $A_{\mathrm{SIDIS}}$ is shown as a function of $x$, $z$, $M_{hh}\equiv M_h$, for the deuterium target ($^6$LiD, upper plot) and for the proton target (NH$_3$, lower plot). No significant asymmetries are observed for the deuterium in any of the variables, suggesting that an effective cancellation is active between the dominant valence up and down contributions because of the isospin symmetry between the proton and neutron components. As for the proton target, the results are consistent with the \texttt{HERMES} findings of Fig.~\ref{fig::hermes_results} after correcting for the depolarization factor $B(y) / A(y)$ in Eq.~(\ref{eq::SIDISssa}) and for a negative sign due to a choice opposite to the Trento conventions. In the \texttt{COMPASS} kinematics, the explored range in $x$ is larger than for the \texttt{HERMES} setup. The lower panel in Fig.~\ref{fig::compass_results_2012} shows a strong dependence of $A_{\mathrm{SIDIS}}$ on $x$, which is directly related to the $x$ dependence of transversity, as displayed by Eq.~(\ref{eq::SIDISssa}).
Recently, a new high-precision measurement on a NH$_3$ target has been published by the \texttt{COMPASS} Collaboration \cite{compass_2014}, that increased the statistics of the first measurement by a factor of four. The new results are in good agreement with the ones discussed above, and provide further constraints on proton transversity.

%
%

The extraction of transversity from the $x$ dependence of the target-spin asymmetry $A_{\mathrm{SIDIS}}$ in Eq.~(\ref{eq::SIDISssa}) implies determining the unknown DiFFs from a different source. Until this was accomplished using the \texttt{BELLE} data for $e^+ e^-$ annihilation (see next section), predictions for $A_{\mathrm{SIDIS}}$ were possible only using model calculations of DiFFs. In Fig.~\ref{fig::compass_results_2012}, the solid lines show an example based on a previously released calculation of DiFFs in the spectator model~\cite{Bacchetta:2006un}, and on the transversity distribution extracted from the analysis of the Collins effect in single-hadron fragmentation~\cite{Anselmino:2008jk}. The dashed lines refer to an analysis based on the pQCD counting rule~\cite{she_2008}.


\begin{figure}[h]
\begin{center}
\resizebox{0.45\textwidth}{!}{%
  \includegraphics{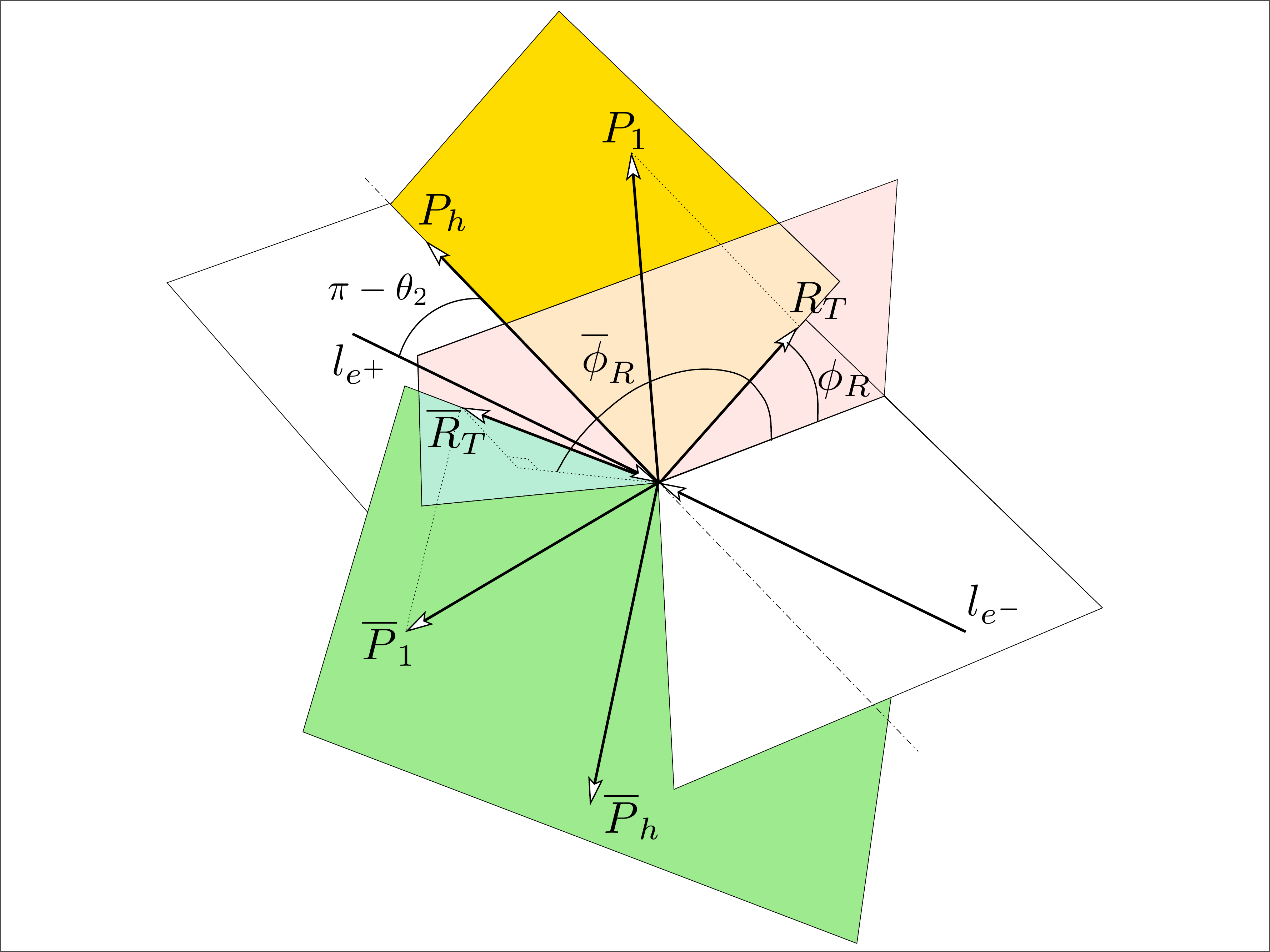}
}
\caption{Definition of the kinematics for the process $e^+ e^- \to (h_1 h_2)_{\mathrm{jet1}} (\bar{h}_1 \bar{h}_2)_{\mathrm{jet2}} X$ where no definition of a thrust axis is needed.}
\label{fig::e+e-kin}
\end{center}
\end{figure}

\vspace{-0.5cm}
\subsection{The Artru-Collins asymmetry in $e^+ e^-$ annihilation}
\label{sec::e+e-ssa}
The main goal is to obtain a model independent information on DiFFs. Similarly to the case of the Collins function, the DiFFs can be independently extracted from electron-positron annihilation producing two pairs of hadrons in opposite hemispheres. The kinematics of the process $e^+ e^- \to (h_1 h_2)_{\mathrm{jet1}} (\bar{h}_1 \bar{h}_2)_{\mathrm{jet2}} X$ is depicted in Fig.~\ref{fig::e+e-kin} with the so-called no-thrust-axis method. An electron and a positron with momenta $\ell_{e^-}$ and $\ell_{e^+}$, respectively, annihilate producing a virtual photon with time-like momentum transfer $q=l_{e^-}+l_{e^+}$, i.e. $q^2=Q^2 \geq 0$. A quark and an antiquark are then emitted, each one fragmenting into a residual jet and a $(h_1 h_2)$ pair with momenta and masses $P_1, M_1,$ and $P_2, M_2,$ respectively ($\bar{P}_1, \bar{M}_1,$ and $\bar{P}_2, \bar{M}_2,$ for the antiquark jet, respectively, and similarly for all other observables pertaining the antiquark hemisphere). The two hadron pairs belong to two jets that are emitted back-to-back, hence they must be detected in opposite emispheres; this condition is granted by requiring that $P_h\cdot \bar{P}_h \approx Q^2$. In Fig.~\ref{fig::e+e-kin}, the lepton frame is identified by the annihilation direction of $\bm{\ell}_{e^+}$ and the axis $\hat{\bm{z}} = \bm{P}_h$, in analogy to the Trento conventions~\cite{Bacchetta:2004jz}. The relative angle is defined as $\theta_2 = \arccos (\bm{\ell_{e^+}}\cdot\bm{P}_h / (|\bm{\ell_{e^+}}|\,|\bm{P}_h|))$ and is related, in the lepton cm frame, to the invariant $y = P_h\cdot \ell_{e^-} / P_h \cdot q$ by $y = (1+\cos\theta_2)/2$. As usual, the azimuthal angles $\phi_R^{}$ and $\bar{\phi}_R^{}$ give the orientation of the planes containing the momenta of the hadron pairs with respect to the lepton frame. They are defined by~\cite{Bacchetta:2008wb,Courtoy:2012ry}
\begin{equation}
\phi_R^{} = 
\frac{(\bm{\ell}_{e^+}\times \bm{P}_h)\,\cdot \bm{R}_{T}}
     {|(\bm{\ell}_{e^+}\times \bm{P}_h)\,\cdot \bm{R}_{T}|} 
\arccos \left( 
        \frac{\bm{\ell}_{e^+}\times \bm{P}_h}{|\bm{\ell}_{e^+}\times \bm{P}_h|}
	\cdot 
	\frac{\bm{R}_{T}\times \bm{P}_h}{|\bm{R}_{T}\times \bm{P}_h|} 
	\right) \, , 
\label{eq::az_angles}
\end{equation}
and similarly for $\bar{\phi}_R$ with $\bm{R}_T \leftrightarrow \bar{\bm{R}}_T$. An alternative kinematical picture can be set up by constructing the thrust axis of the two back-to-back jets and identifying it with the $\hat{\bm{z}}$ axis (thrust-axis method). In this frame, both pairs' total momenta have non-vanishing transverse components $\bm{P}_{hT}$ and $\bar{\bm{P}}_{hT}$. It has been checked that the final experimental results are quite stable against the choice of the two different methods~\cite{Vossen:2011fk}.

In the framework of collinear factorization, i.e. after integrating upon all transverse momenta but $\bm{R}_{T}$ and $\bar{\bm{R}}_{T}$, the leading-twist cross section for the production of two unpolarized hadron pairs can be written as~\cite{Boer:2003ya,Courtoy:2012ry}
\begin{equation}
\begin{split}
&\frac{d\sigma}{dy\, dz\,dM_h\, d\cos\theta\,d\phi_R^{}\, d\bar{z}\, d\bar{M}_h\, d\cos\bar{\theta}\, d\bar{\phi}_R^{}\, dQ^2} = \\
&\mbox{\hspace{1cm}} \frac{1}{4\pi^2}\, d\sigma^0\, \left[ 1 + \cos (\phi_R^{}+\bar{\phi}_R^{})\, A_{e^+e^-} \right] \; , 
\label{eq::e+e-cross}
\end{split}
\end{equation}
where
\begin{equation}
\begin{split}
&\frac{d\sigma^0}{dy\, dz\,dM_h\,d\bar{z}\, d\bar{M}_h\, dQ^2} = \frac{3\pi \alpha^2}{2 Q^2} \, \frac{\langle 1+\cos^2\theta_2\rangle}{4} \\
&\qquad \times \sum_q e_q^2\, D_1^q (z, M_h^2; Q^2)\, \bar{D}_1^q (\bar{z}, \bar{M}_h^2; Q^2)  
\label{eq::e+e-cross0}
\end{split}
\end{equation}
is the unpolarized part and 
\begin{equation} 
\begin{split} 
&A_{e^+e^-}  =  \frac{\sin^2 \theta_2}{ 1+\cos^2 \theta_2} \, \sin\theta \sin\bar{\theta} \, \frac{|\bm{R}|}{M_h} \, \frac{|\bar{\bm{R}}|}{\bar{M}_h} \\
&\mbox{\hspace{.5cm}} \times \, \frac{\sum_q e_q^2 \, H_1^{\sphericalangle\, q}(z, M_h^2; Q^2)\, \bar{H}_1^{\sphericalangle \, q}(\bar{z}, \bar{M}_h^2; Q^2)}
      {\sum_q e_q^2\, D_1^q (z, M_h^2; Q^2) \, \bar{D}_1^q (\bar{z}, \bar{M}_h^2; Q^2)} 
\label{eq::e+e-ssa}
\end{split} 
\end{equation} 
is the so-called Artru-Collins asymmetry~\cite{Boer:2003ya}. In all formulas above, the flavor sum is understood to run over quarks and antiquarks. 

The non-perturbative correlation of Eq.~(\ref{eq::H1angle_dens}) is responsible for the azimuthally asymmetric term $A_{e^+e^-}$ in Eq.~(\ref{eq::e+e-cross}). If the back-to-back jets were produced by unpolarized quark-antiquark pairs, the distribution of detected hadron pairs would be azimuthally symmetric and would show the peculiar dependence $1+\cos^2\theta_2$, as it is the case in the unpolarized cross section $d\sigma^0$. The presence of the $\cos (\phi_R + \bar{\phi}_R) \, \sin^2 \theta_2$ modulation points out that also a  transversely polarized $q^\uparrow \bar{q}^\downarrow$ pair is produced from the $e^+e^-$ annihilation, each parton fragmenting into a pair of hadrons in its own jet. The transverse polarization then is correlated to the asymmetric orientation of the planes containing the momenta of the two hadron pairs, the correlation being described by the IFF $H_1^{\sphericalangle \, q}$ for the involved flavor $q$. 


\begin{figure}[h]
\begin{center}
\resizebox{0.37\textwidth}{!}{%
  \includegraphics{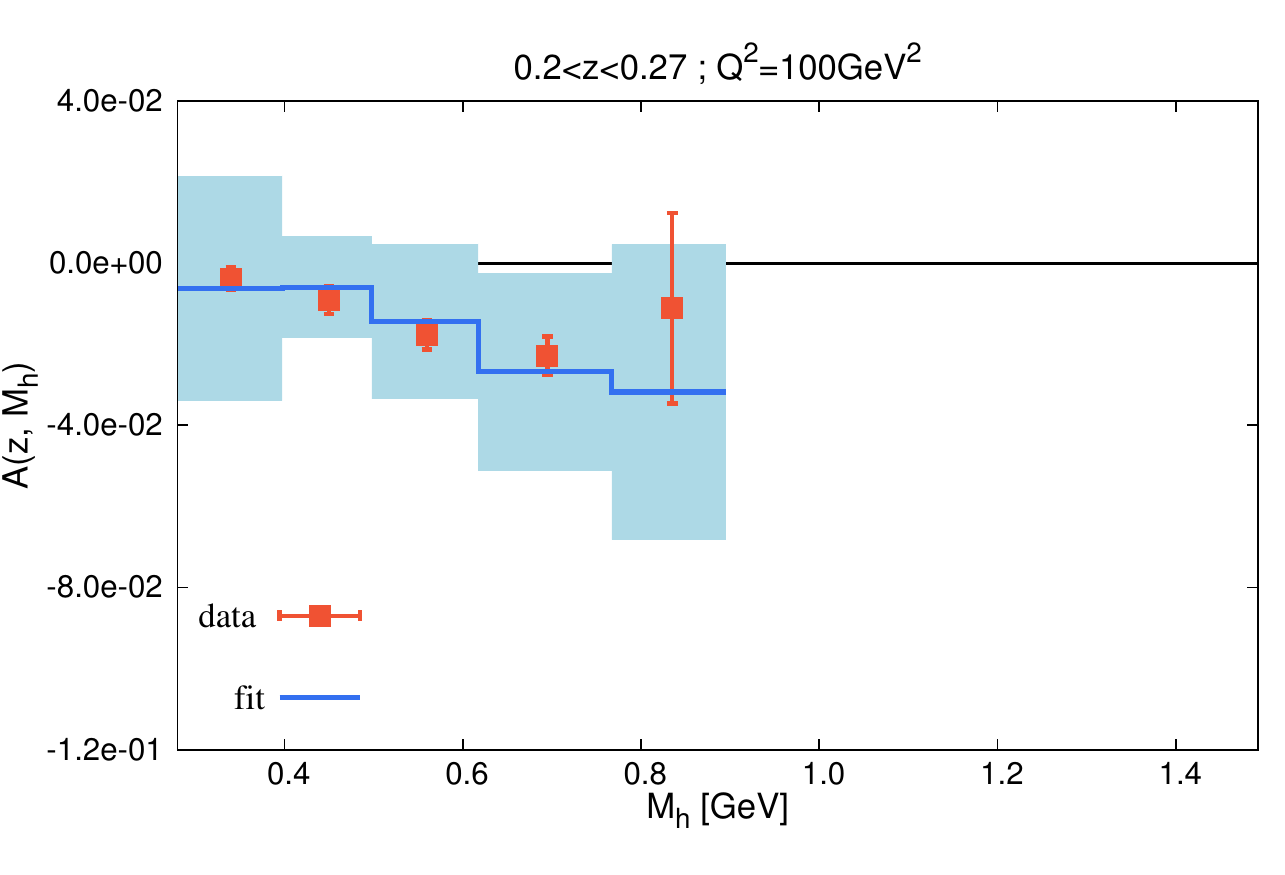}}\\
\resizebox{0.37\textwidth}{!}{%
   \includegraphics{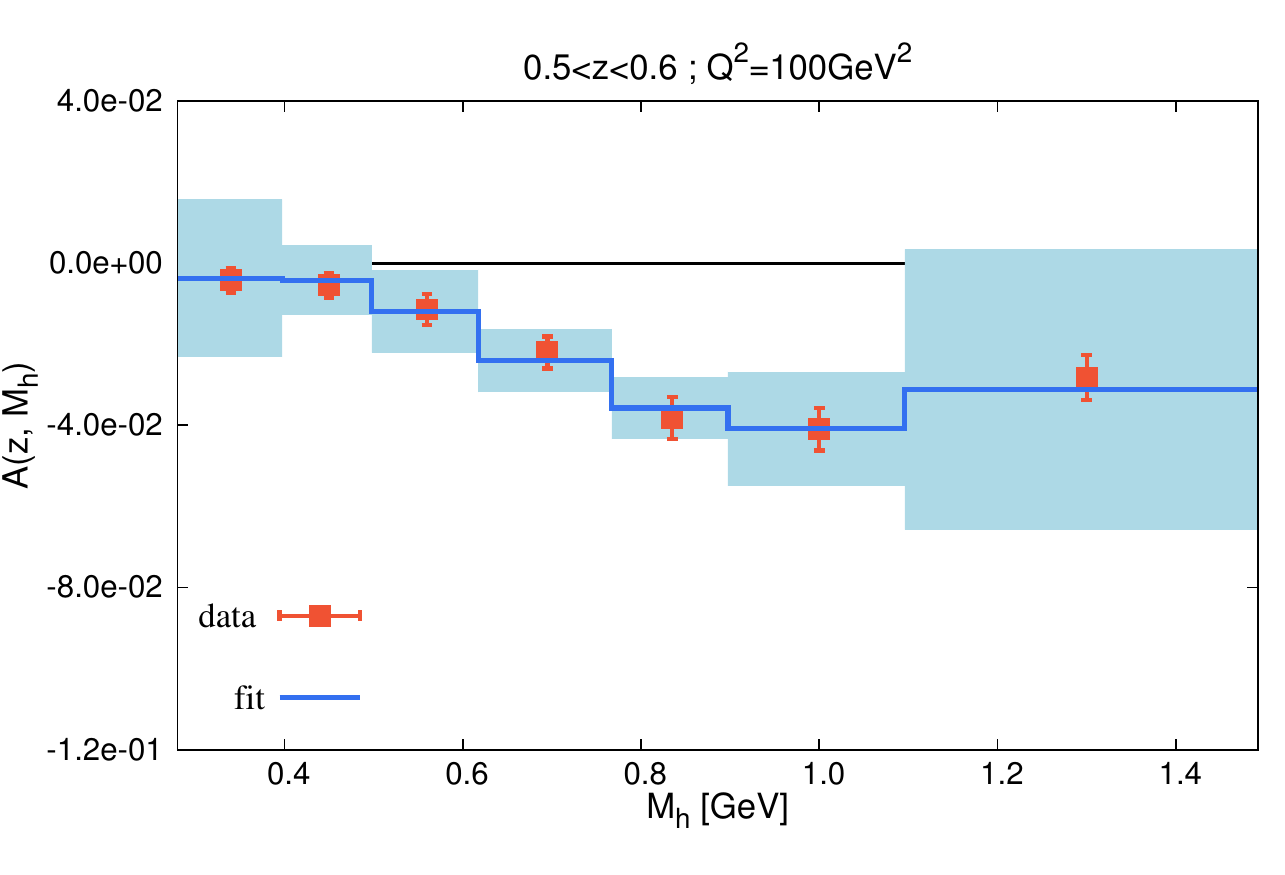}
}
\caption{Histogram of the Artru-Collins asymmetry at $Q^2=100$ GeV$^2$ in bins of  invariant mass $M_h$ for the $z$ bins $[0.2, 0.27]$ (upper panel) and $[0.5, 0.6]$ (lower panel)~\cite{Courtoy:2012ry}. Experimental points with error bars from \texttt{BELLE}~\cite{Vossen:2011fk}. Solid line represents the top side of the fitting histogram. Shaded area is the corresponding statistical error.}
\label{fig::e+e-ssa}
\end{center}
\end{figure}

\vspace{-1cm}
\subsection{The BELLE experiment: the extraction of DiFFs}
\label{sec::BELLE}

The $\cos (\phi_R + \bar{\phi}_R)\, \sin^2\theta_2$ modulation was predicted in Ref.~\cite{Boer:2003ya} and measured for the first time by the \texttt{BELLE} collaboration~\cite{Vossen:2011fk} for the case of detected $\pi^+ \pi^-$ pairs. The data sample was collected at the KEKB energy-asymmetric collider operating at a cm energy around the ${\cal Y}(4S)$ resonance. All pions were required to have a minimal fractional energy $z > 0.1$ in order to count only pairs coming from a genuine $q\to (\pi \pi)$ fragmentation. By summing over all pairs detected in one emisphere, data for the Artru-Collins asymmetry $A_{e^+e^-}$ were collected in a $8\times 8$ grid of $(z, M_h)$ bins. In Fig.~\ref{fig::e+e-ssa}, they are shown as bullets with statistical error bars for the $z$ bins $[0.2, 0.27]$ (upper panel) and $[0.5, 0.6]$ (lower panel). 

In the same figure, the solid line indicates the top side of the fitting histogram, where the shaded area is its statistical error. The fitting formula is derived from Eq.~(\ref{eq::e+e-ssa}) with some further manipulation. First of all, both the unpolarized and polarized parts of the cross section must be integrated, e.g., over $\bar{z}$ and $\bar{M}_h$ within the experimental cuts, in order to account for all the pairs in one emisphere. Then, the flavor sum is simplified because DiFFs are symmetric under isospin transformation and charge conjugation when the final hadrons are a $(\pi^+ \pi^-)$ pair~\cite{Courtoy:2012ry,Bacchetta:2006un,Bacchetta:2011ip}. Thus, the Artru-Collins asymmetry simplifies to
\begin{equation}
\begin{split} 
A_{e^+e^-} & = - \frac{\sin^2 \theta_2}{1+\cos^2 \theta_2} \, \sin\theta \sin\bar{\theta} \, \frac{5}{9} \\
&\times \frac{|\bm{R}|}{M_h} \, 
\frac{H_1^{\sphericalangle \, u} (z, M_h^2; Q^2)\, n_u^\uparrow (Q^2)}
      {\sum_q e_q^2\, D_1^q (z, M_h^2; Q^2) \, n_q (Q^2)} \; ,
\label{eq::e+e-ssafit}
\end{split} 
\end{equation} 
where 
\begin{equation}
\begin{split}
n_q(Q^2) &= \int_{0.2}^1 dz \int_{2 m_\pi}^2 dM_h \, D_1^q (z, M_h^2; Q^2)   \\
n_q^\uparrow (Q^2) &= \int_{0.2}^1 dz \int_{2 m_\pi}^2 dM_h \, \frac{|\bm{R}|}{M_h}\, H_1^{\sphericalangle\, q}(z,M_h^2; Q^2) \; ,
\label{eq::nDiFF}
\end{split}
\end{equation}
and the flavor sum in the denominator is limited to the lightest four flavors. 

In order to extract $H_1^{\sphericalangle \, u}$ from $A_{e^+e^-}$, one needs to know the unpolarized DiFF $D_1$ first. Contrary to the case of single-hadron fragmentation, no data are available yet for the unpolarized cross section for the semi-inclusive production of $(\pi^+ \pi^-)$ pairs. Therefore, in Ref.~\cite{Courtoy:2012ry} $D_1$ was parametrized to reproduce the two-pion yield of the \texttt{PYTHIA} event generator tuned to the \texttt{BELLE} kinematics. The fitting expression at the starting scale $Q_0^2=1$ GeV$^2$ was inspired by previous model calculations~\cite{Bacchetta:2008wb,Radici:2001na,Bacchetta:2006un,Bianconi:1999uc} and it contains three resonant channels (pion pair produced by $\rho$, $\omega$, and $K^0_S$ decays) and a continuum. For each channel and for each flavor $q= u,d,s,c$, a grid of data in $(z,M_h)$ was produced using \texttt{PYTHIA} for a total amount of approximately 32000 bins. Each grid was separately fitted using the corresponding parametrization of $D_1$ and evolving it to the \texttt{BELLE} scale at $Q^2=100$ GeV$^2$. An average $\chi^2$ per degree of freedom ($\chi^2$/d.o.f.) of 1.62 was reached using in total 79 parameters (see  Ref.~\cite{Courtoy:2012ry} for further details). 

As for the polarized DiFF, it is convenient to manipulate Eq.~(\ref{eq::e+e-ssafit}) and define the following function
\begin{equation} 
\begin{split}
H(z, &M_h; Q^2) \equiv \frac{|\bm{R}|}{M_h} \, H_{1}^{\sphericalangle u}(z, M_h; Q^2)\, n_u^\uparrow (Q^2) \\
&= - \frac{1+\cos^2 \theta_2}{\sin^2 \theta_2}\, \frac{9}{5} \,\frac{1}{\sin\theta \, \sin\bar{\theta}} \, A_{e^+e^-} \\
&\mbox{\hspace{1.5cm}} \times \sum_q e_q^2\, D_1^q (z, M_h^2; Q^2) \, n_q (Q^2) \; ,
\label{eq::ssafitnum}
\end{split}
\end{equation} 
with the normalization
\begin{equation}
\int dz \int dM_h \, H(z, M_h, Q^2) = [ n_u^\uparrow (Q^2)]^2 \; . 
\label{eq::Hnorm}
\end{equation}

%
\begin{figure}[h]
\begin{center}
\resizebox{0.35\textwidth}{!}{%
  \includegraphics{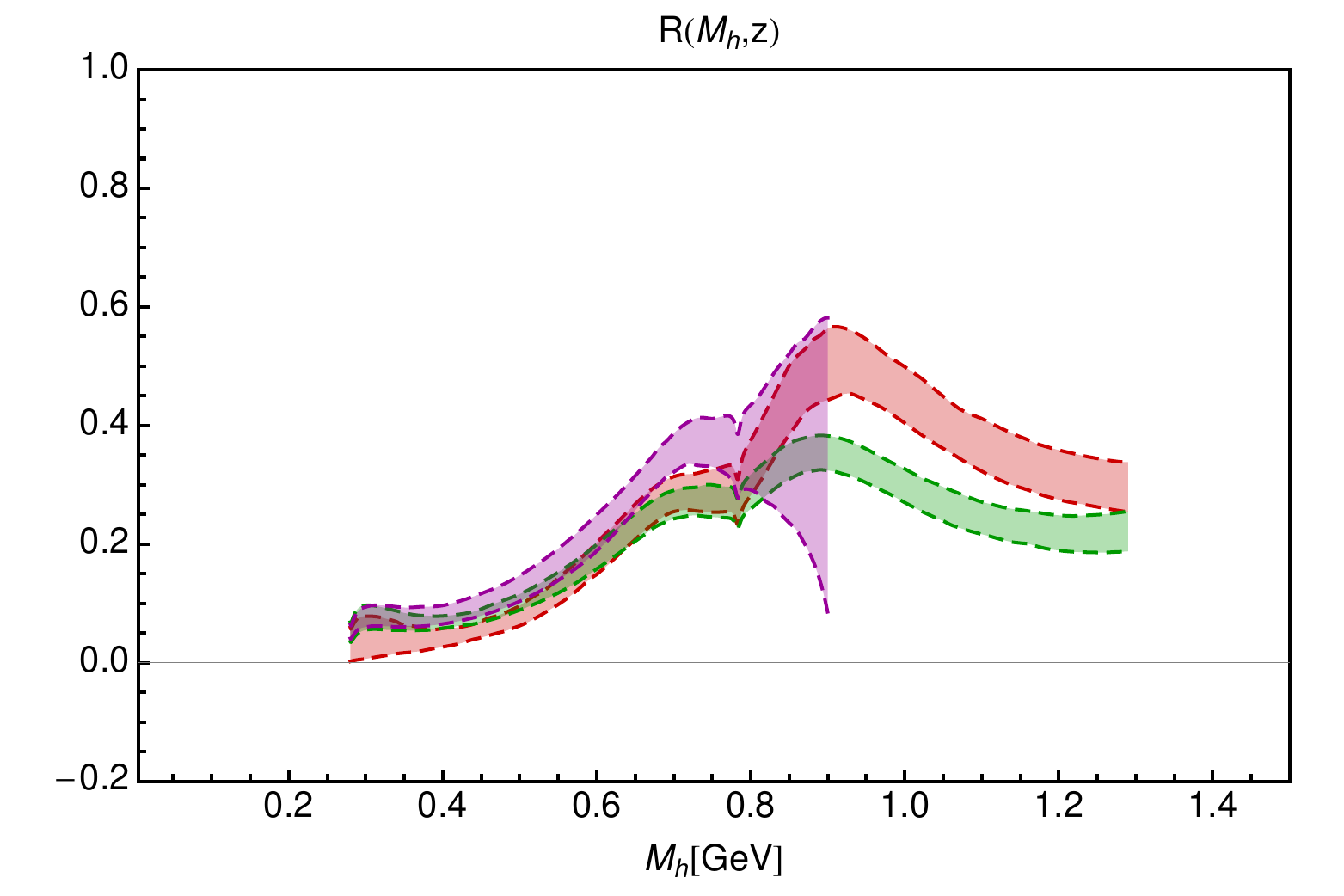}}\\
\resizebox{0.35\textwidth}{!}{%
   \includegraphics{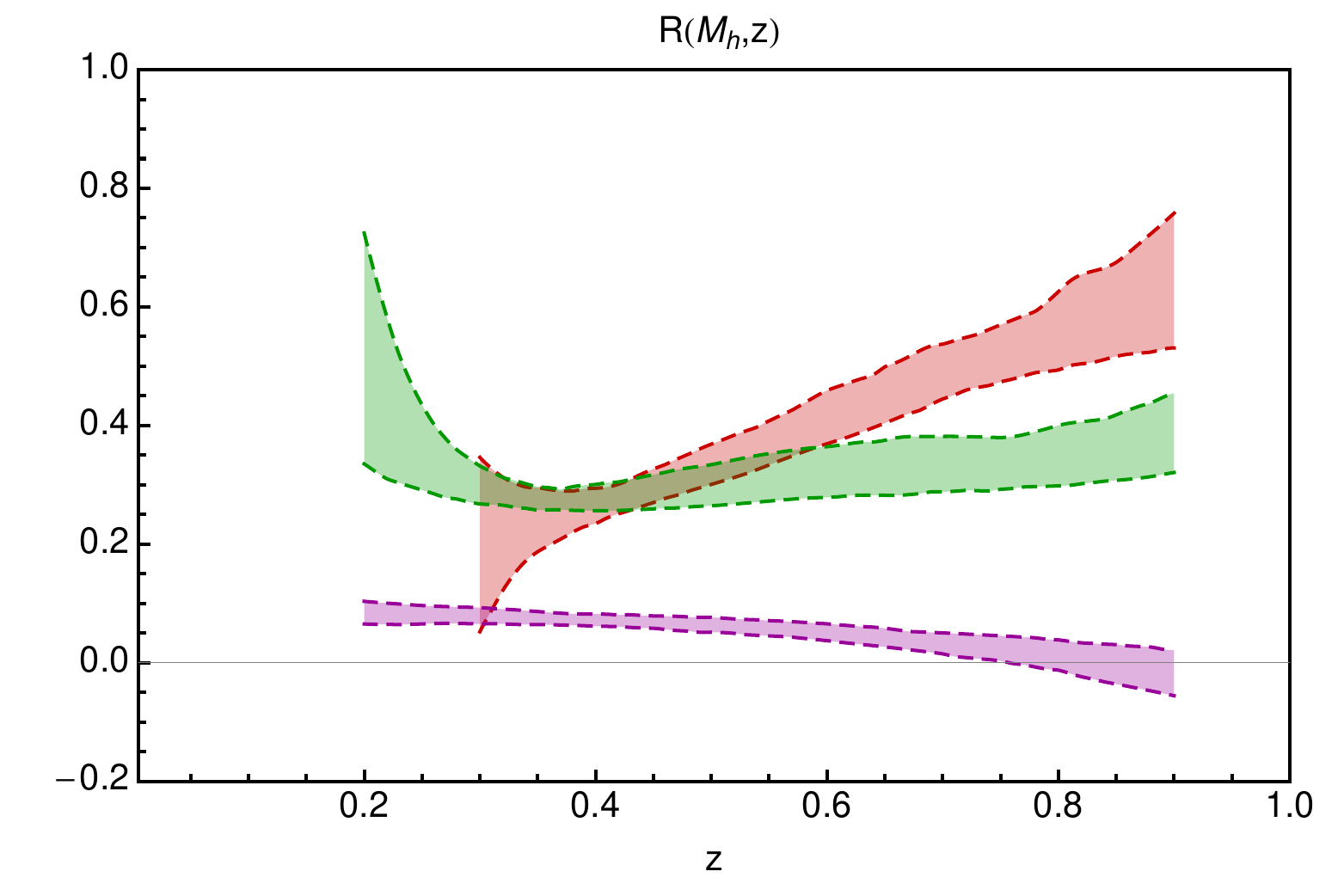}
}
\caption{The ratio $|\bm{R}|/M_h\  H_1^{\sphericalangle\, u}/D_1^u$ at $Q_0^2=1$ GeV$^2$~\cite{Radici:2015mwa}. Upper panel: ratio as function of $M_h$ for $z=0.25$ (shortest band), $z=0.45$ (lower band at $M_h \sim 1.2$ GeV), and $z=0.65$ (upper band at $M_h \sim 1.2$ GeV). Lower panel: ratio as function of $z$ for $M_h=0.4$ GeV (lower band at $z\sim 0.8$), $M_h=0.8$ GeV (mid band at $z\sim 0.8$), and $M_h=1$ GeV (upper band at $z\sim 0.8$). }
\label{fig::H1uD1u}
\end{center}
\end{figure}
%

In each bin, the experimental value of $H$ in Eq.~(\ref{eq::ssafitnum}) is deduced from the experimental data for the Artru-Collins asymmetry $A_{e^+e^-}$ (organized in a $8\times 8$ grid of $(z, M_h)$ bins) and the corresponding average values of the angles $\theta_2, \, \theta, \, \bar{\theta}$ (all taken from Ref.~\cite{Vossen:2011fk}), and from the unpolarized DiFFs $D_1^q$ resulting from the fit of the \texttt{PYTHIA}'s two-pion yield. The fitting formula for $H$ at the starting scale $Q_0^2=1$ GeV$^2$ depends on 9 parameters~\cite{Courtoy:2012ry}. It is then evolved to the \texttt{BELLE} scale of $Q^2 = 100$ GeV$^2$ and tuned to its experimental value. The error analysis of the first extraction was performed with the traditional Hessian method, reaching a very good $\chi^2$/d.o.f. = 0.57 and producing the fitting histograms in Fig.~\ref{fig::e+e-ssa}~\cite{Courtoy:2012ry}. Then, the analysis was repeated in Ref.~\cite{Radici:2015mwa} using a different approach, which consists in perturbing the experimental points with a Gaussian noise to create $M$ replicas of them, and in separately fitting the $M$ replicas. The final outcome is a set of $M$ different values of the vector of 9 fitting parameters or, equivalently, of $M$ different fitting functions $H$. These $M$ values are not necessarily distributed with a Gaussian shape; the 68\% uncertainty band can be simply obtained by rejecting the largest and smallest 16\% of values for each experimental bin. The value of $M$ is determined by accurately reproducing the mean and standard deviation of the original data points. The replica method is more general and, consequently, more reliable because it does not rely on the prerequisites for the standard Hessian method to be valid; the latter are often not fulfilled when the minimization procedure pushes the fitting functions towards the boundaries of the phase space (see Ref.~\cite{Radici:2015mwa} for further details). 

In Fig.~\ref{fig::H1uD1u}, the ratio $|\bm{R}|/M_h\  H_1^{\sphericalangle \, u}/D_1^u$ is shown as a function of $M_h$ (upper panel) and of $z$ (lower panel) at $Q_0^2=1$ GeV$^2$~\cite{Radici:2015mwa}. The various bands represent the 68\% of the $M=100$ replicas obtained, case by case, with the procedure explained above. In the upper panel, the shortest band corresponds to $z=0.25$, the lower band at $M_h \sim 1.2$ GeV to $z=0.45$, and the upper band at $M_h \sim 1.2$ GeV to $z=0.65$, respectively. In all cases, the peaks corresponding to the $\rho$ and $\omega$ resonances are clearly visible. In the lower panel, the lower band at $z\sim 0.8$ corresponds to $M_h=0.4$ GeV, the mid band at $z\sim 0.8$ to $M_h=0.8$ GeV, the upper band at $z\sim 0.8$ to $M_h=1$ GeV.

\subsection{The extraction of transversity}
\label{sec::h1}

The \texttt{BELLE} measurement of the Artru-Collins asymmetry $A_{e^+e^-}$~\cite{Vossen:2011fk}, and the following parametrization of DiFFs~\cite{Courtoy:2012ry,Radici:2015mwa}, represent a turning point because they have made possible the extraction of transversity in a collinear framework using Eq.~(\ref{eq::SIDISssa}) in a model independent way. As before, for the $(\pi^+\pi^-)$ case the symmetry properties of DiFFs under isospin transformations and charge conjugation~\cite{Bacchetta:2006un,Bacchetta:2011ip} simplify the flavor sum in Eq.~(\ref{eq::SIDISssa}). Moreover, the $x$-dependence of transversity is more conveniently studied by integrating the $z$- and $M_h$-dependences of DiFFs. 

The analysis of the \texttt{HERMES} data for the target-spin asymmetry $A^p_{\text{SIDIS}}$ for a transversely polarized proton target (see Sec.~\ref{sec::HERMES}) gives access to the following combination~\cite{Bacchetta:2011ip}:
\begin{equation} 
\begin{split}
x\, &h_1^p (x; Q^2) \equiv x \, h_1^{u_v}(x; Q^2) - {\textstyle \frac{1}{4}}\, x h_1^{d_v}(x; Q^2) \\
&= -\frac{ A^p_{\text{SIDIS}} (x; Q^2)  }{n_u^{\uparrow}(Q^2)} \, \frac{A(y)}{B(y)}\, \frac{9}{4}\, \sum_q e_q^2 \, n_q (Q^2)\, x f_1^{q+\bar{q}}(x; Q^2)  \; ,
\label{eq::xh1p} 
\end{split}   
\end{equation} 
where $h_1^{q_v} \equiv h_1^q - h_1^{\bar{q}}$ and $f_1^{q+\bar{q}} \equiv f_1^q + f_1^{\bar{q}}$. Using a common parametrization for $f_1(x)$ (for example, the \texttt{MSTW08} set of Ref.~\cite{Martin:2009iq}) and the \texttt{HERMES} data for the target-spin asymmetry $A^p_{\text{SIDIS}}$~\cite{hermes_2008}, all the unknowns in the right-hand side of Eq.~(\ref{eq::xh1p}) are determined because the $n_u^{\uparrow}(Q^2)$ and $n_q (Q^2)$ can be computed for $q=u,d,s,$ at the $Q^2$ of each \texttt{HERMES} data point from the extracted DiFFs and from their evolution equations~\cite{Ceccopieri:2007ip}. In Ref.~\cite{Bacchetta:2011ip}, the first point-by-point extraction of $x h_1^p$ was performed in this way and compared with the corresponding expression built on the transversity extracted from the Collins effect; the agreement was reasonable, although the small number of experimental points did not allow to draw any conclusion. 


\begin{figure}[h]
\begin{center}
\resizebox{0.45\textwidth}{!}{%
  \includegraphics{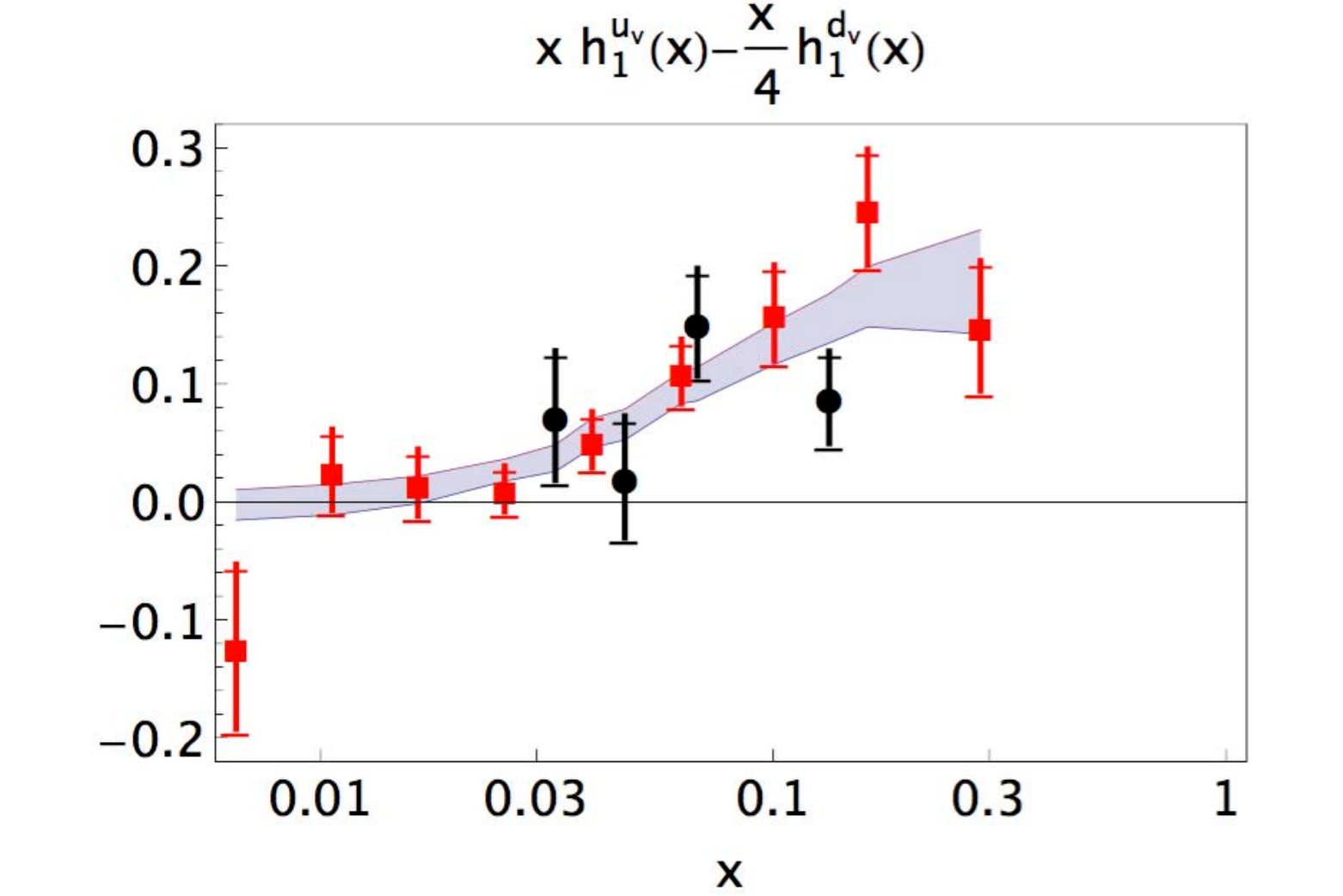}}\\
\resizebox{0.45\textwidth}{!}{%
   \includegraphics{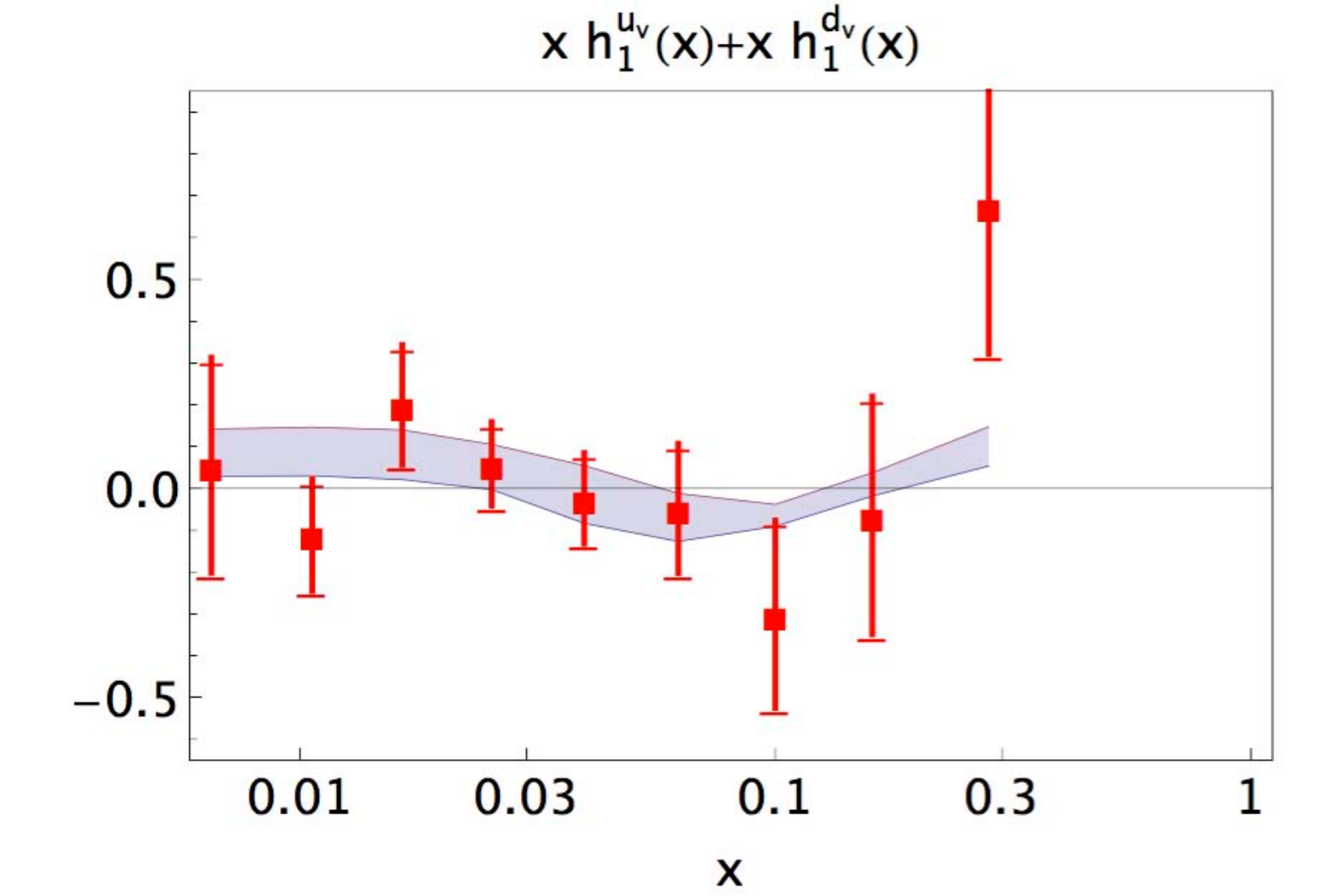}
}
\caption{The combinations of Eq.~(\ref{eq::xh1p}) (upper panel) and Eq.~(\ref{eq::xh1D}) (lower panel). The black circles are obtained from the \texttt{HERMES} data for the SSA $A^p_{\text{SIDIS}}$~\cite{hermes_2008}; the lighter squares from the \texttt{COMPASS} data for both $A^p_{\text{SIDIS}}$~\cite{Braun:2015baa} and $A^D_{\text{SIDIS}}$~\cite{compass_2012}. The uncertainty band represents the selected $68\%$ of all fitting replicas in the rigid scenario with $\alpha_s (M_Z^2)=0.125$ (see text).}
\label{fig::SIDISxh1p+D}
\end{center}
\end{figure}


When the \texttt{COMPASS} results for the target-spin asymmetry became available~\cite{compass_2012}, the analysis included also data for a transversely polarized deuteron target. These data can be used in a flavor combination independent from the one in Eq.~(\ref{eq::xh1p}), namely~\cite{Bacchetta:2012ty,Radici:2015mwa}
\begin{equation} 
\begin{split} 
 x\, &h_1^D (x; Q^2) \equiv x \, h_1^{u_v}(x; Q^2)+ x h_1^{d_v}(x; Q^2)   \\
 &=- \frac{A^D_{\text{SIDIS}}(x; Q^2)}{n_u^{\uparrow}(Q^2)} \, \frac{A(y)}{B(y)}\, 3 \\
 &\quad \  \times \sum_q \, \left[ e_q^2 \, n_q (Q^2) + e^2_{\tilde{q}} \, n_{\tilde{q}} (Q^2) \right] \, x f_1^{q+\bar{q}}(x; Q^2) \; ,
\label{eq::xh1D}
\end{split} 
\end{equation} 
where $\tilde{q} = d,u,s$ if $q = u,d,s,$ respectively ({\it i.e.}, it reflects isospin symmetry of strong interactions inside the deuteron). In Ref.~\cite{Bacchetta:2012ty}, the point-by-point extraction of $x h_1^D$ was made using the \texttt{COMPASS} data for the deuteron target in $A^D_{\text{SIDIS}}$ from the 2004 run, and the point-by-point extraction of $x h_1^p$ was improved by adding to the \texttt{HERMES} data also the \texttt{COMPASS} data for the proton target in $A^p_{\text{SIDIS}}$ from the 2007 run for unidentified $(h^+ h^-)$ pairs~\cite{compass_2012}. In Ref.~\cite{Radici:2015mwa}, the same analysis was repeated by inserting in $A^p_{\text{SIDIS}}$ the most recent and more precise \texttt{COMPASS} data for identified $(\pi^+\pi^-)$ pairs produced off proton targets from the 2010 run~\cite{Braun:2015baa}. In Fig.~\ref{fig::SIDISxh1p+D}, the point-by-point extractions of $x h_1^p$ and $x h_1^D$ are shown in the upper and lower panels, respectively. The black squares in the upper panels correspond to the \texttt{HERMES} data for $A^p_{\text{SIDIS}}$~\cite{hermes_2008}, all the other points refer to the \texttt{COMPASS} data of Ref.~\cite{Braun:2015baa} (upper panel) and of Ref.~\cite{compass_2012} (lower panel). The error bars are obtained by propagating the statistical errors in the formulas of Eqs.~(\ref{eq::xh1p}) and~(\ref{eq::xh1D}); they are dominated by the experimental errors on $A^p_{\text{SIDIS}}$ and $A^D_{\text{SIDIS}}$, respectively. 

By combining $x h_1^p$ and $x h_1^D$, the valence components of transversity can be separated point-by-point~\cite{Elia:2012}. If we further fit the experimental values for $x h_1^p$ and $x h_1^D$ shown in Fig.~\ref{fig::SIDISxh1p+D}, we can get a separate parametrization of the $x$-dependence of each valence flavor of transversity at a given $Q^2$ scale. The strategy is similar to the one adopted for extracting the polarized DiFF $H_1^{\sphericalangle}$ by fitting the experimental values for the function $H$ in Eq.~(\ref{eq::ssafitnum}). Namely, the $A^p_{\text{SIDIS}}$ and $A^D_{\text{SIDIS}}$ data are perturbed with a Gaussian noise in order to create $M$ replicas of them that are separately fitted. The fitting function is built in order to make transversity satisfy the Soffer's inequality at any scale~\cite{Bacchetta:2012ty,Radici:2015mwa}. The final outcome is again a set of $M$ different values of the vector of fitting parameters, and the 68\% uncertainty band is formed by rejecting the largest and smallest 16\% of the $M$ results for each experimental bin. Three different scenarios have been explored, depending of the number of parameters: at the starting scale the fitting function contains a polynomial in $x$ that can have 1 node ("rigid" scenario), 2 nodes ("flexible" scenario), or 3 nodes ("extraflexible" scenario). The function is then evolved to the $Q^2$ scale of each data point, using different values for the normalization of the strong coupling constant at the $Z$ boson mass $(\alpha_s (M_Z^2))$ in order to account for the theoretical uncertainty in determining the $\Lambda_{\text{QCD}}$ parameter (for more details, see Refs.~\cite{Bacchetta:2012ty,Radici:2015mwa}). In  Ref.~\cite{Bacchetta:2012ty}, the analysis was performed using the \texttt{COMPASS} data of Ref.~\cite{compass_2012}; in Ref.~\cite{Radici:2015mwa}, the analysis was updated using the most recent \texttt{COMPASS} data for proton targets of Ref.~\cite{Braun:2015baa}. In Fig.~\ref{fig::SIDISxh1p+D}, the uncertainty bands display the 68\% of $M=100$ replicas computed in Ref.~\cite{Radici:2015mwa} in the rigid scenario using $\alpha_s (M_Z^2) = 0.125$~\cite{Gluck:1998xa}. 


\begin{figure}[h]
\begin{center}
\resizebox{0.45\textwidth}{!}{%
  \includegraphics{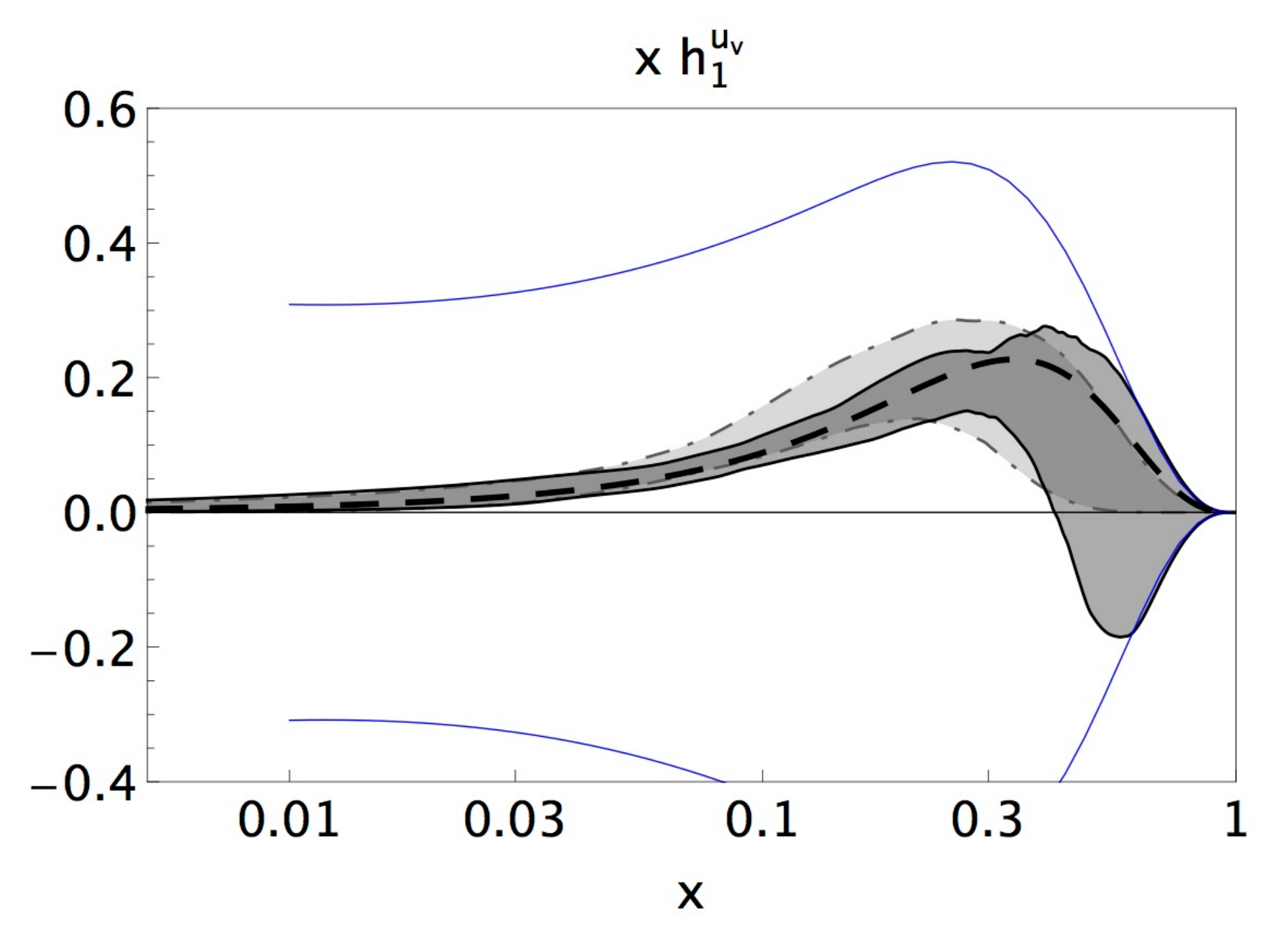}}\\
\resizebox{0.45\textwidth}{!}{%
   \includegraphics{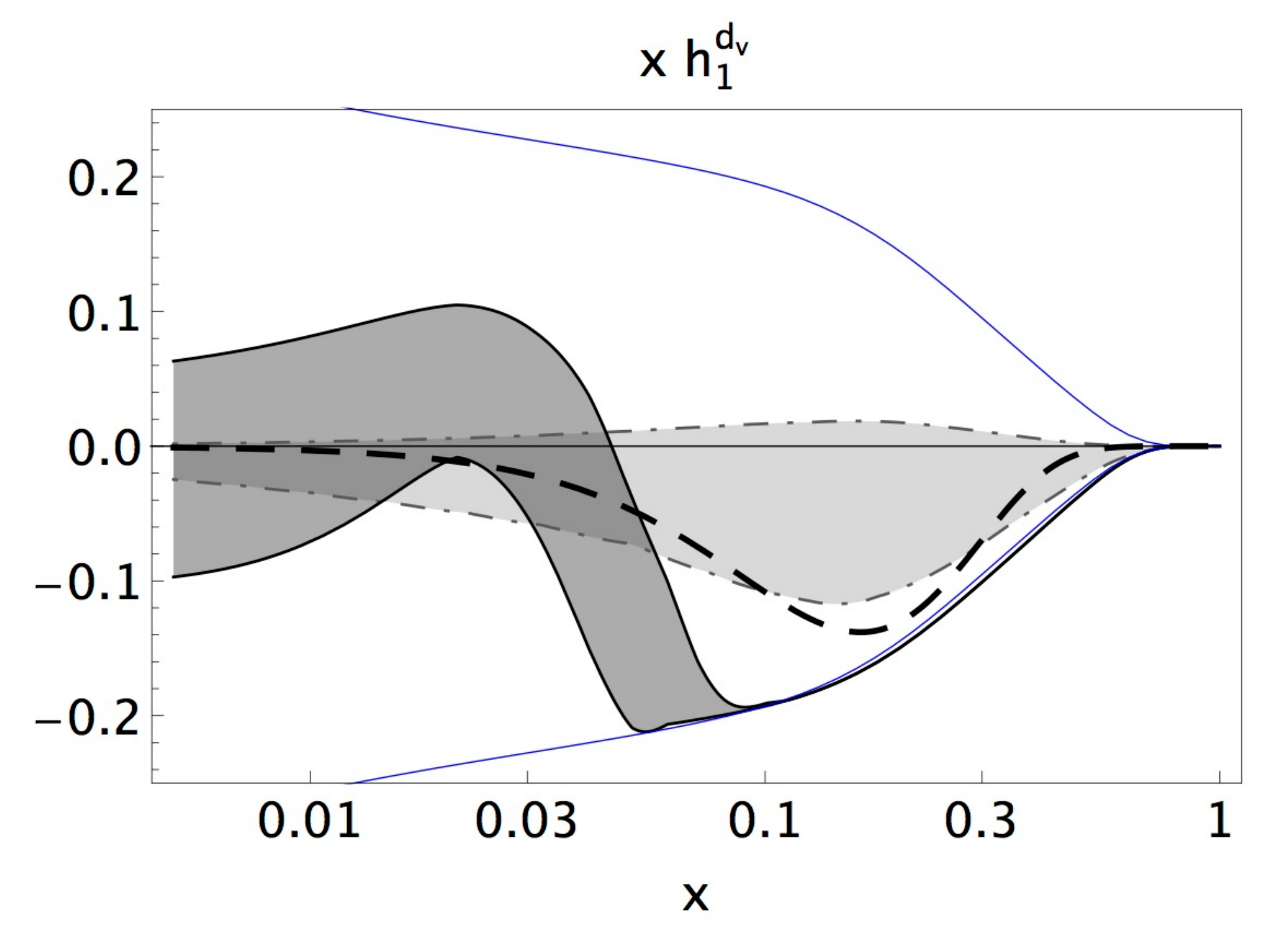}
}
\caption{The up (upper panel) and down (lower panel) valence transversities as functions of $x$ at $Q^2=2.4$ GeV$^2$. The darker band with solid borders is the result of ref.~\cite{Radici:2015mwa} in the flexible scenario with $\alpha_s (M_Z^2)=0.125$. The lighter band with dot-dashed borders is the most recent transversity extraction from the Collins effect~\cite{Anselmino:2013vqa}. The central thick dashed line is the result of Ref.~\cite{Kang:2014zza}. The thick solid lines indicate the Soffer bound.}
\label{fig::SIDISxh1u+d}
\end{center}
\end{figure}


In Fig.~\ref{fig::SIDISxh1u+d}, the dark bands with solid borders show the corresponding valence transversities (up quark in the upper panel, down quark in the lower panel) as functions of $x$ at $Q^2 = 2.4$ GeV$^2$ in the flexible scenario and for $\alpha_s (M_Z^2) = 0.125$~\cite{Radici:2015mwa}. The thick solid lines indicate the Soffer bound, which represents the uncrossable border for the replicas by construction. The lighter band with dot-dashed borders in the background is the most recent transversity extraction of Ref.~\cite{Anselmino:2013vqa} using the Collins effect but applying the standard DGLAP evolution equations only to the collinear part of the fitting function. The central thick dashed line is the result of Ref.~\cite{Kang:2014zza}, where evolution equations have been computed in the TMD framework. The latter analysis has been recently updated~\cite{Kang:2015msa} including also a calculation of the error band which turns out to mostly overlap with the lighter band from Ref.~\cite{Anselmino:2013vqa}. There is a general consistency among the various extractions, at least for the range $0.0065 \leq x \leq 0.29$ where there are data. This remark has been confirmed by the recent analysis of Ref.~\cite{Martin:2014wua} where the method of point-by-point extraction has been extended to the case of single-hadron SIDIS, and the transversity distributions obtained with the two different mechanisms (di-hadron production or Collins effect) have been shown to be compatible with each other. This is encouraging: despite the fact that the dihadron SIDIS data are a subset of the single-hadron ones (in Ref.~\cite{Martin:2014wua}, it is even argued that they are two different manifestations of the same mechanism), the theoretical frameworks used to interpret them are very different. Nevertheless, we point out that the collinear framework, in which results with DiFFs are produced, represents a well established and robust theoretical context. On the contrary, the implementation of the QCD evolution equations of TMDs, particularly for chiral-odd functions like the transversity and the Collins function, is not yet fully settled~\cite{Kang:2015msa}. Moreover, the error analysis based on the replica method gives a more realistic description of the uncertainty on transversity, specifically for large $x$ outside the data range. As it is clear in the upper panel of Fig.~\ref{fig::SIDISxh1u+d}, for $x \geq 0.3$ the replicas tend to fill all the phase space available within the Soffer bound. In order to reduce this uncertainty, it is important that new data will be collected in this region with the forthcoming upgrade of Jefferson Lab to the 12 GeV beam. In the lower panel, the discrepancy between the results around $x \geq 0.1$ is confirmed in all the scenarios explored: rigid, flexible, or extraflexible. It can be ascribed to the behavior of two specific bins of the \texttt{COMPASS} data for the deuteron target, where the values of $A^D_{\text{SIDIS}}$ drive the replicas to saturate the Soffer bound~\cite{Bacchetta:2012ty,Radici:2015mwa}. It is also interesting to remark that the dashed line from Ref.~\cite{Kang:2014zza}, although in general agreement with the other extraction based on the Collins effect, also tends to saturate the Soffer bound at $x > 0.2$. 

The first Mellin moment of transversity for a flavor $q$ gives the tensor charge $\delta q$. The similarity of the parametrized transversities in Fig.~\ref{fig::SIDISxh1u+d} reflects in compatible results also for the tensor charges: when performing the integral over the whole $x$ range the extrapolation outside the data range increases the uncertainty and smooths the differences~\cite{Radici:2015mwa,Kang:2015msa}. Also the calculated isovector tensor charge $g_T = \delta u_v - \delta d_v$ is in agreement with many lattice calculations~\cite{Radici:2015mwa}. The $g_T$ belongs to the group of isovector nucleon charges that are related to flavour-changing processes. A determination of these couplings may shed light on the search of new physics mechanisms that may depend on 
them~\cite{Bhattacharya:2011qm,Ivanov:2012qe,Cirigliano:2013xha,Courtoy:2015haa}, or on direct dark matter searches~\cite{DelNobile:2013sia}.


\begin{figure}[h]
\begin{center}
\resizebox{0.45\textwidth}{!}{%
  \includegraphics{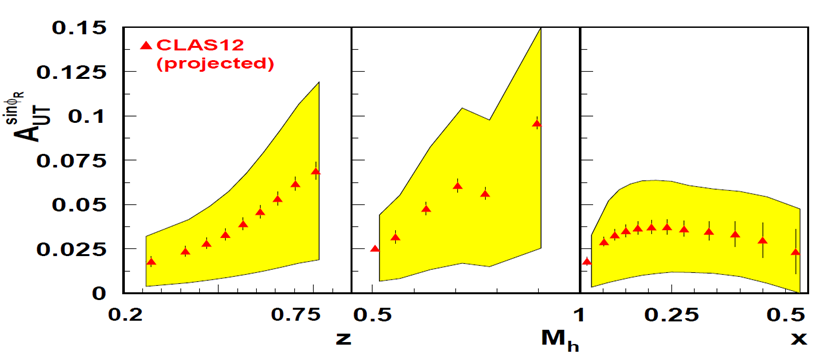}
}\\
\resizebox{0.5\textwidth}{!}{%
   \includegraphics{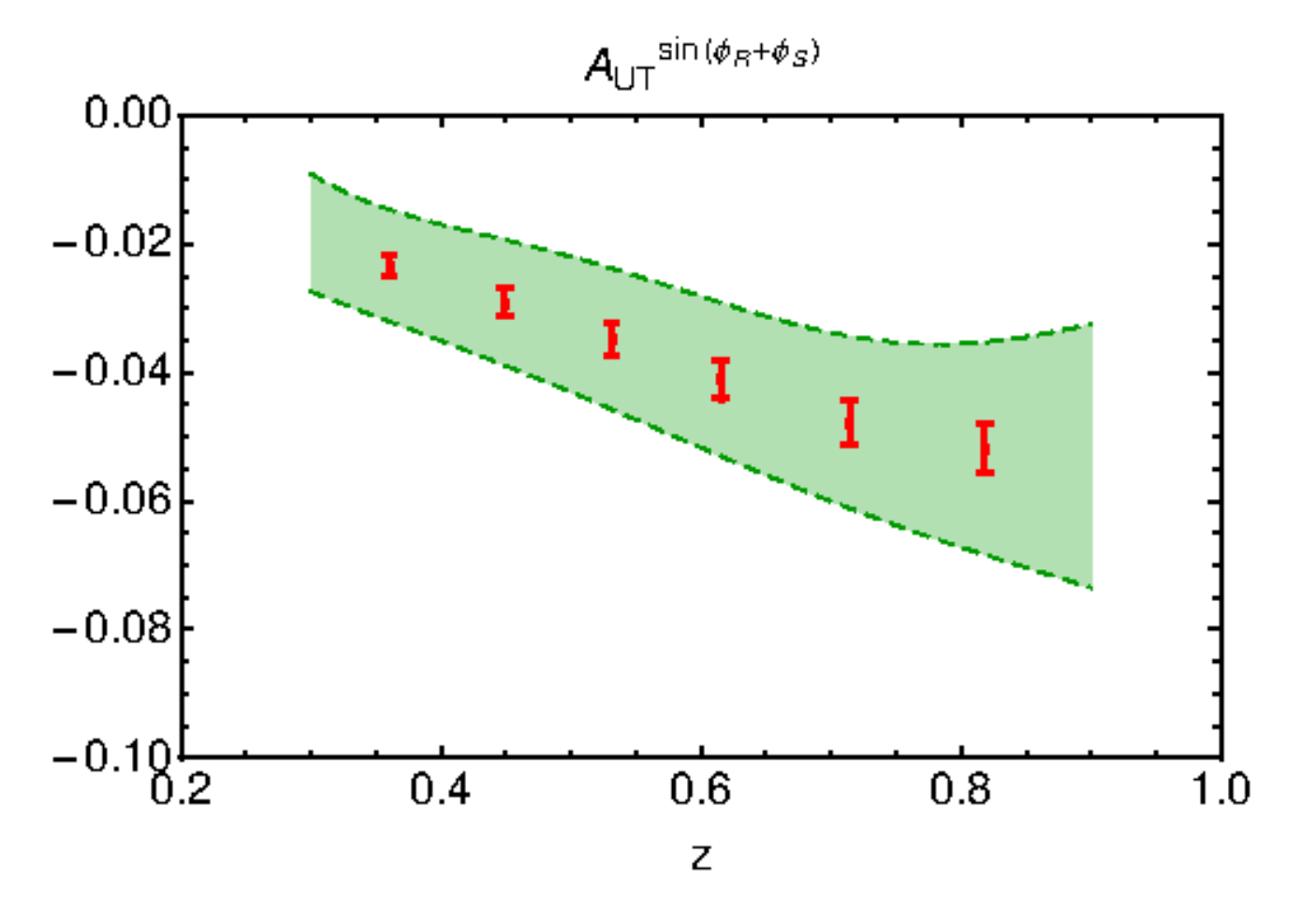} \includegraphics{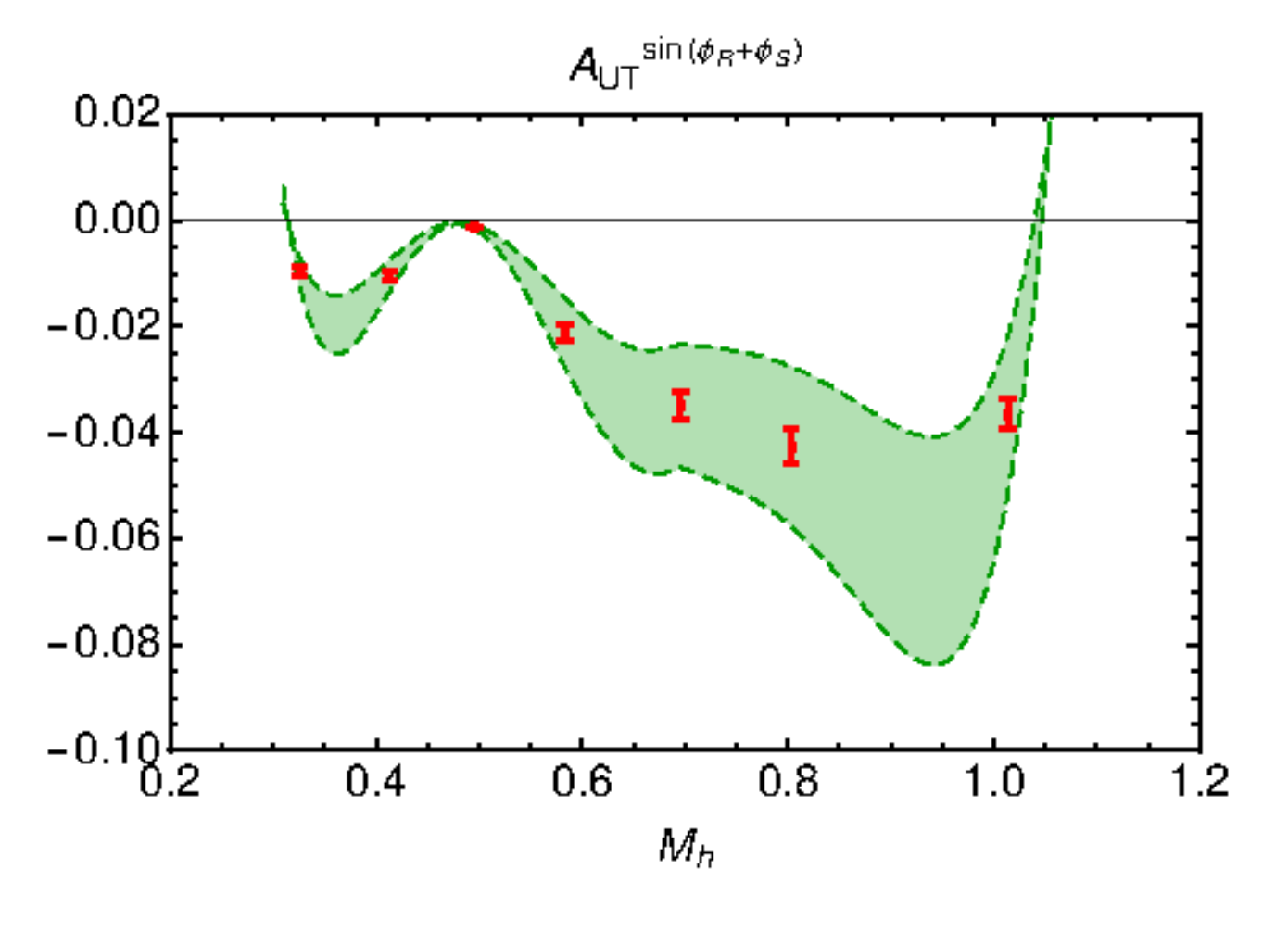} \includegraphics{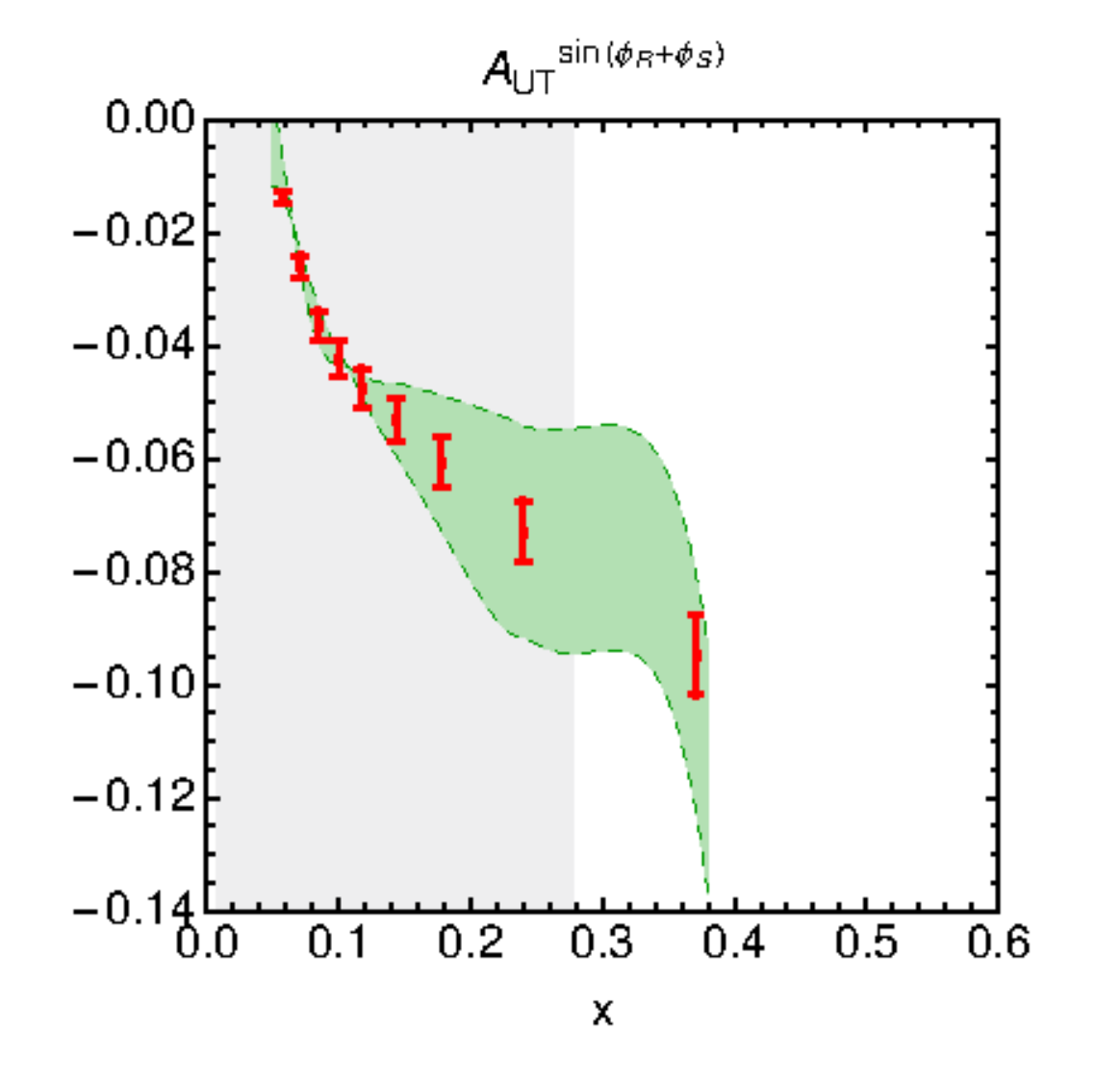}
}
\caption{The target-spin asymmetry in SIDIS as a function of $z$ (left panels), $M_h$ (middle panels) and $x$ (right panels). Upper panels: projections for the \texttt{CLAS12} detector in Hall B on a transversely polarized proton target. Lower panels: projections for the \texttt{SoLID} detector in Hall A on a transversely polarized $^3$He target. The band in the lower right panel represents the region in $x$ covered by the present \texttt{HERMES} and \texttt{COMPASS} data.}
\label{fig::dh_clas12_solid}
\end{center}
\end{figure}


\subsection{Future measurements at JLab in the 11-GeV era}
\label{sec::jlab}

The Continuous Electron Beam Accelerator Facility (CEBAF) at Jefferson Lab is undergoing an upgrade~\cite{jlab12_upgrade} that will double the energy of the electrons provided to the four experimental halls operating in the 12-GeV era (Hall A, B, C, D). Hall B will be equipped with a large-acceptance spectrometer (\texttt{CLAS12}~\cite{clas12_detector}). A proposal relative to the measurement of di-hadron transverse observables on a tranversely polarized proton target has been conditionally approved~\cite{dh_clas12_p}. Also in Hall A the possibility to build a large-acceptance solenoid spectrometer (\texttt{SoLID}) is being explored, and different physics proposals have been already approved that make use of this new device~\cite{solid_sidis,solid_12_11_007,solid_12_11_108}. In particular, the transverse target-spin asymmetry $A_{\text{SIDIS}}$ will be measured on a transversely polarized $^3$He target, providing access to the neutron transversity distribution in the valence region~\cite{dh_solid_n}. The high statistics provided by \texttt{CLAS12} and \texttt{SoLID} measurements will allow for a multidimensional binning of $A_{\text{SIDIS}}$. 

In Fig.~\ref{fig::dh_clas12_solid}, the projections for $A_{\text{SIDIS}}$ as a function of $z$ (left panels), $M_h$ (middle panels), and $x$ (right panels), are shown. The upper panels display the expectations for the \texttt{CLAS12} detector using a transversely polarized proton target. The lower panels are related to the \texttt{SoLID} detector adopting a transversely polarized $^3$He nucleus as an effective transversely polarized neutron target. In all panels, the bands give a measure of the overall uncertainty on the transversity distribution, obtained by merging the uncertainties in the extraction of $h_1$ from the analysis of the Collins effect or of the DiFF effect (compare with Fig.~\ref{fig::SIDISxh1u+d} in the previous section). The band in the lower, right panel represents the region in $x$ covered by the present \texttt{HERMES} and \texttt{COMPASS} data. Thus, the two measurements at the Jefferson Lab will provide an excellent coverage in the valence region. Due to the overlapping kinematics, the results of the two experiments will eventually be combined to separately access the $u$ and $d$ flavors of transversity.

\subsection{The extraction of unpolarized DiFF}
\label{sec::d1}

\begin{figure}[h]
\begin{center}
\resizebox{0.45\textwidth}{!}{%
   \includegraphics{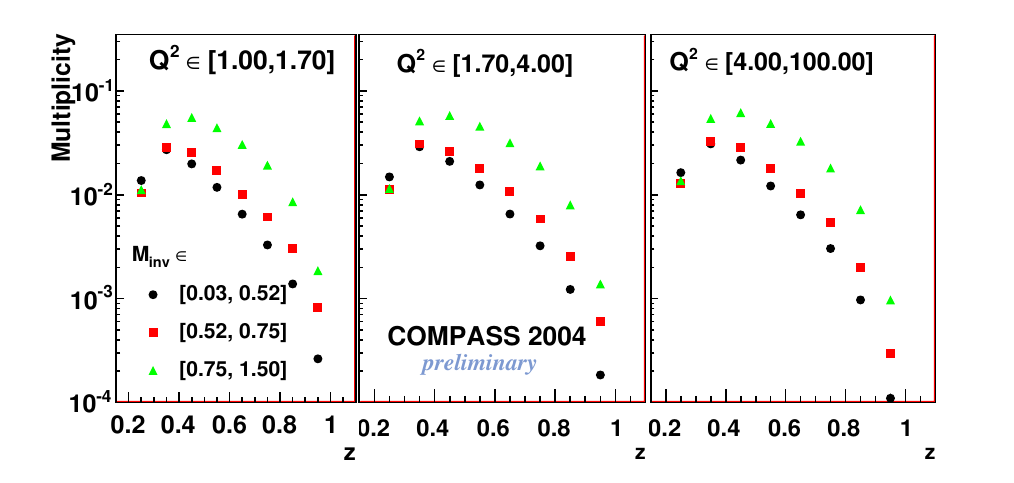}
}
\caption{The SIDIS multiplicity for the semi-inclusive production of a $(\pi^+ \pi^-)$ pair at \texttt{COMPASS} kinematics~\cite{Makke:2013dya}.}
\label{fig::multi_compass}
\end{center}
\end{figure}

As already stressed in Sec.~\ref{sec::BELLE}, the unpolarized DiFF $D_1$ was extracted by fitting the output of the \texttt{PYTHIA} Monte Carlo adapted to the \texttt{BELLE} kinematics, because no data are available yet for the unpolarized cross section in $e^+e^-$ annihilations. Another useful observable is the SIDIS multiplicity
\begin{eqnarray}
M(x, z, M_h^2; Q^2) &= &\frac{d\sigma / dx\, dz\, dM_h^2\, dQ^2}{d\sigma^{\text{incl}} / dx\, dQ^2} \nn \\
&\propto &\frac{\sum_q e_q^2 \, f_1^q (x; Q^2)\, D_1^q (z, M_h^2; Q^2)}{\sum_q e_q^2\, f_1^q(x; Q^2)} \; , 
\label{eq::DiFFmulti}
\end{eqnarray}
where $d\sigma$ is the cross section for the $ep \to e' (h^+, h^-) X$ process and $d\sigma^{\text{incl}}$ is the corresponding inclusive one. Work is in progress by the \texttt{COMPASS} Collaboration. In Fig.~\ref{fig::multi_compass}, the preliminary results for $M$ are shown as functions of $z$ for various bins in $M_h$ and $Q^2$~\cite{Makke:2013dya}. 


\begin{figure}[h]
\begin{center}
\resizebox{0.35\textwidth}{!}{%
   \includegraphics{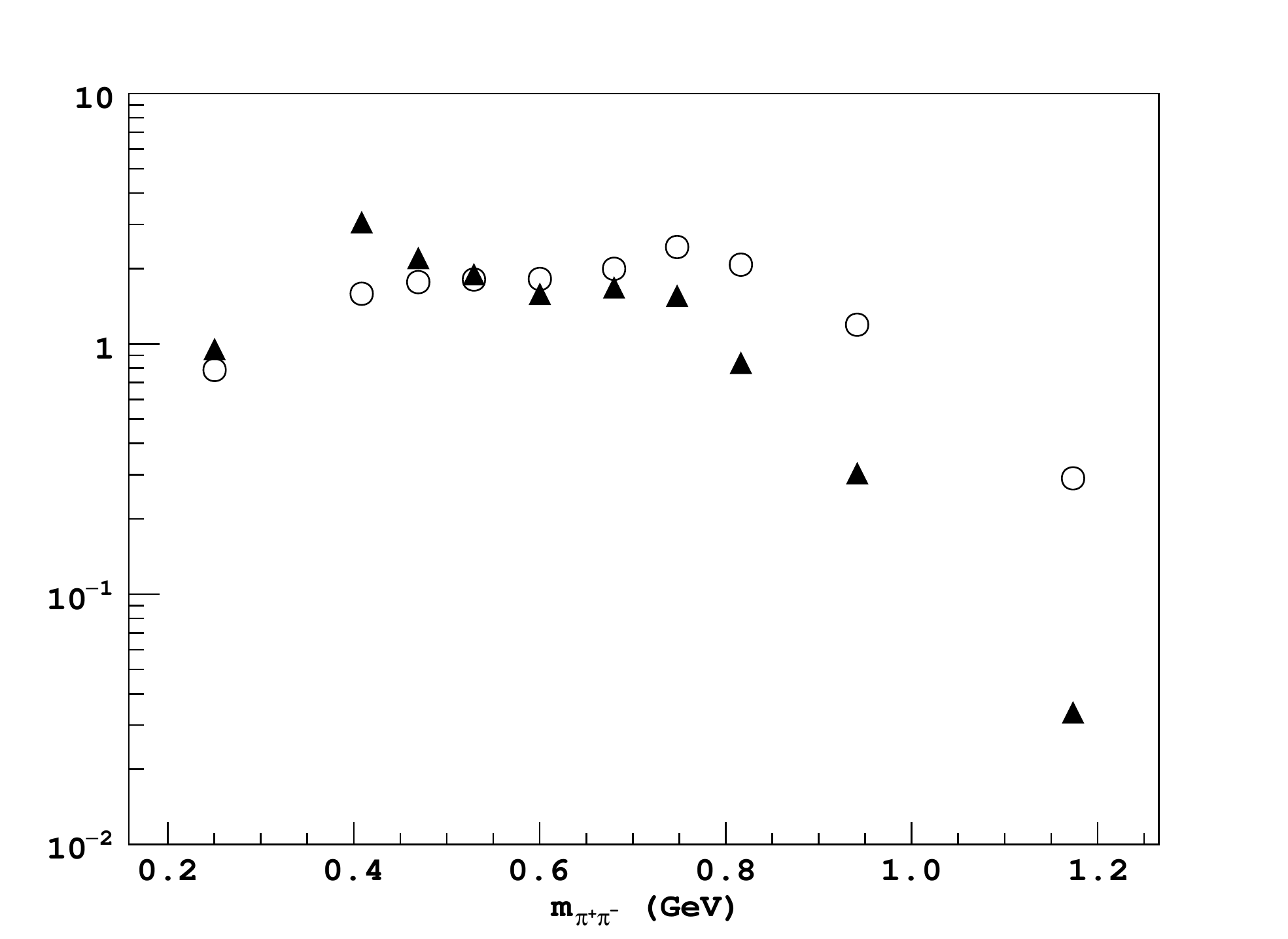}
}
\caption{Solid triangles: the SIDIS multiplicity $M$ of Eq.~(\ref{eq::DiFFmulti}) for the $(\pi^+ \pi^-)$ production off a proton target in bins of $M_h$ for a $(Q^2;\, x,\, z)$ bin, as extracted from \texttt{PEPSI} generated data. Empty circles: the same observable built by combining the $D_1^q$ from the unpolarized cross section for the $e^+ e^- \to (\pi^+ \pi^-) X$ process at \texttt{BELLE} (as generated from \texttt{PYTHIA}~\cite{Courtoy:2012ry}) with the PDF $f_1^q$ from the \texttt{MSTW08LO} set.}
\label{fig::multi_clas12}
\end{center}
\end{figure}


Another measurement of $M$ for $(\pi^+ \pi^-)$ pairs has been recently approved~\cite{prop_clas_11GeV} in the \texttt{CLAS12} experiment~\cite{clas12_detector} that will be installed in Hall B at Jefferson Lab~\cite{jlab12_upgrade}. Both unpolarized hydrogen and deuterium targets will be employed, allowing for the flavor separation of the $u$ and $d$ components of $D_1$. A multidimensional binning in $(Q^2, x, z, M_h)$ will be performed for a total of $5\times 10\times 10\times 10 = 5000$ kinematical points. In Fig.~\ref{fig::multi_clas12}, an example of the expected precision is shown. Solid triangles refer to the SIDIS multiplicity of Eq.~(\ref{eq::DiFFmulti}) for the $(\pi^+ \pi^-)$ production off a proton target in bins of $M_h$ for a $(Q^2;\, x,\, z)$ bin, as extracted from \texttt{PEPSI} generated data. They are compared with the empty circles that describe the SIDIS multiplicity by combining the $D_1^q$ obtained from \texttt{PYTHIA} (adapted to reproduce at \texttt{BELLE} kinematics the unpolarized cross section for the $e^+ e^- \to (\pi^+ \pi^-) X$ process~\cite{Courtoy:2012ry}) with the PDF $f_1^q$ from the \texttt{MSTW08LO} set. A similar precision is expected for the deuterium data. In both cases, the Monte Carlo simulations are directly proportional to the unpolarized DiFF $D_1$. As already discussed in Sec.~\ref{sec::BELLE}, a better knowledge of $D_1$ would improve the extraction of the polarized DiFFs, in particular of the IFF $H_1^{\sphericalangle}$ and, consequently, of the transversity distribution.

\subsection{Is transversity a universal distribution?}
\label{sec::h1pp}

The agreement displayed in Fig.~\ref{fig::SIDISxh1u+d} among the various extractions of transversity is a first important cross-check about the reliability of the results. But the actual verification of transversity being a universal parton distribution implies that the obtained parametrization $h_1^{q_v} (x, Q^2)$ can be used in different contexts and energies to make predictions for different processes involving transversely polarized partons. To this aim, DiFFs turn again to be useful. 

In fact, let us consider the collision $p p^\uparrow \rightarrow (h_1 h_2) X$ where a proton with momentum $P_A$ collides on a transversely polarized proton with momentum $P_B$ and spin vector $S_B$, producing a pair of unpolarized hadrons $h_1, h_2$ inside the same jet. The transverse component of the total pair momentum $\bm{P}_h$ with respect to the beam $\bm{P}_A$ is indicated with $\bm{P}_{h\perp}$ and serves as the hard scale of the process. If the kinematics is collinear, namely if the transverse component $\bm{P}_{hT}$ of $\bm{P}_h$ around the jet axis is integrated over, the differential cross section at leading order in $1/|\bm{P}_{h\perp}|$, {\it i.e.} at leading twist, is~\cite{Bacchetta:2004it}
\begin{equation}
\frac{d\sigma}{d\eta\, d|\bm{P}_{h\perp}|\, dM_h^2\,d\phi_R} =  
 d\sigma^0 \  \left( 1 + \sin (\phi_{S_B} - \phi^{}_R) \, A_{pp} \right)  
\label{eq::ppcross}
\end{equation}
with $\phi_{S_B} = \pi / 2$, where 
\begin{equation}
\begin{split}
d\sigma^0&=  2 \, |\bm{P}_{h\perp}| \, \sum_{a,b,c,d}\int \frac{d x_a\, dx_b }{4 \pi^2 z_h} \\ &\quad \times f_1^a (x_a) \, f_1^b(x_b) \, \frac{d\hat{\sigma}_{ab \to cd}}{d\hat{t}} \, 
D_1^c (z_h, M_h^2) 
\label{eq::ppcross0}
\end{split}
\end{equation}
and 
\begin{equation}
\begin{split}
A_{pp}&= [d\sigma^0]^{-1}\, 2 \, |\bm{P}_{h\perp}|\, \frac{|\bm{R}|}{M_h}\,|\bm{S}_{BT}|\, \sum_{a,b,c,d}\, 
\int \frac{dx_a \, dx_b}{16 \pi z_h} \\
&\quad \times f_1^a(x_a) \, h_1^b(x_b) \, \frac{d\Delta \hat{\sigma}_{ab^\uparrow \to c^\uparrow d}}{d\hat{t}} H_1^{\sphericalangle c}(z_h, M_h^2) \, .
\label{eq::App}
\end{split}
\end{equation}


\begin{figure}[h]
\begin{center}
\resizebox{0.3\textwidth}{!}{%
  \includegraphics{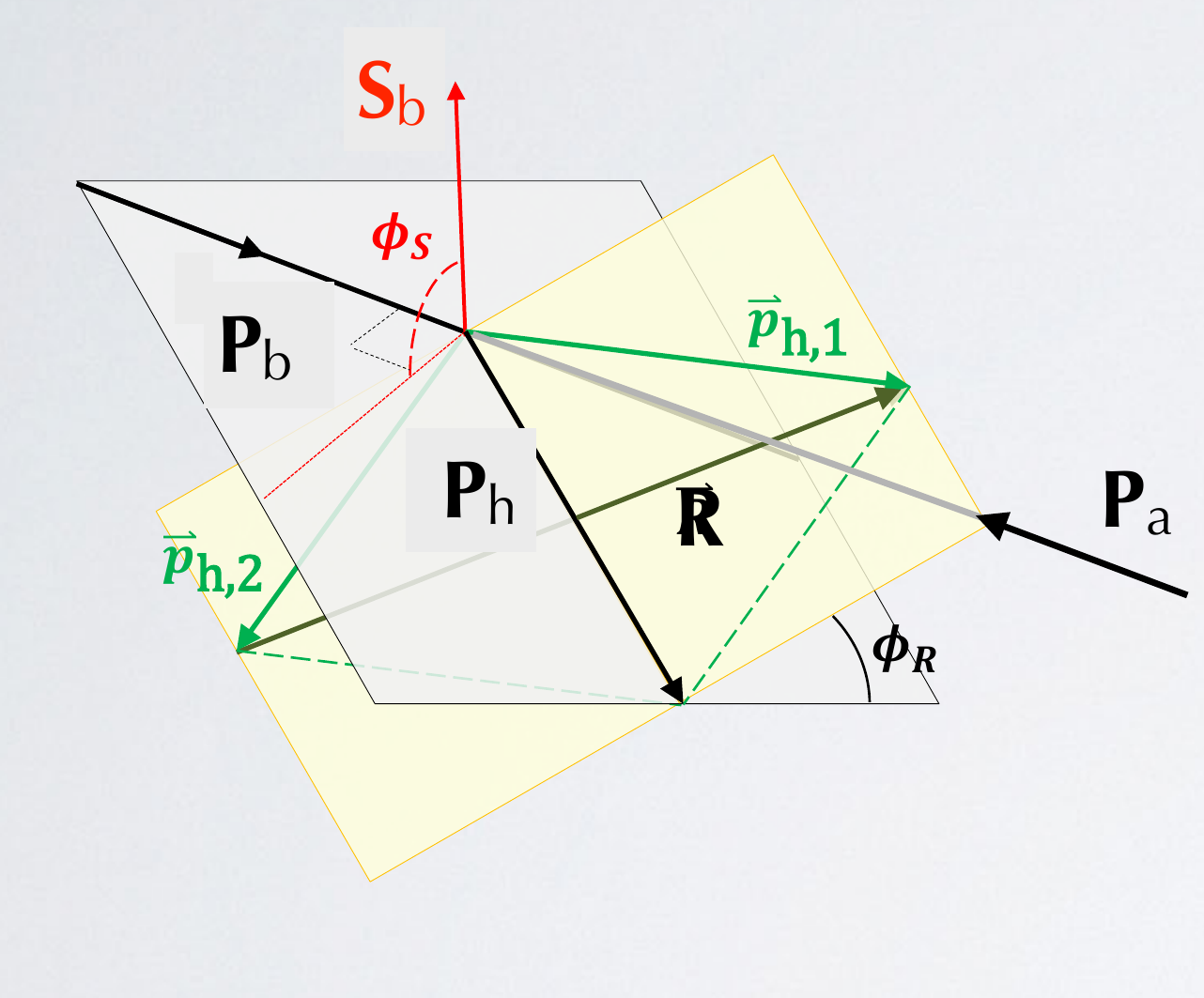}
}
\caption{Kinematics for a proton colliding on a transversely polarized proton and leading to the semi-inclusive production of two hadrons inside the same jet.}
\label{fig::ppDiFFkin}
\end{center}
\end{figure}


In Eq.~(\ref{eq::App}), $\bm{R}$ is the usual relative vector of the final hadron pair (see Fig.~\ref{fig::ppDiFFkin}) and its azimuthal angle is defined as 
\begin{equation}
\begin{split}
\cos \phi_{R} &= 
  \frac{\hat{\bm{P}}_h \times \bm{P}_A}{|\hat{\bm{P}}_h\times\bm{P}_A|}
  \cdot \frac{\bm{R} \times \hat{\bm{P}}_h}{|\bm{R}\times\hat{\bm{P}}_h|} \, , \\
\sin \phi_{R} &= 
  \frac{(\bm{R} \times \bm{P}_A) \cdot \hat{\bm{P}}_h}{|\hat{\bm{P}}_h\times\bm{P}_A|\,|\hat{\bm{P}}_h\times\bm{R}|} \, . 
\label{eq::pp_phiR}
\end{split}
\end{equation}
Moreover, the pseudorapidity $\eta$ is defined by~\cite{Bacchetta:2004it}
\begin{equation}
z_h = \frac{|\bm{P}_{h\perp}|}{\sqrt{s}}\, \frac{x_a e^{-\eta} + x_b e^{\eta}}{x_a x_b} \, , 
\label{eq::rapidity}
\end{equation}
where $\sqrt{s}$ is the cm energy of the collision, and $\hat{t} = t \  x_a / z_h$, with $t = (P_A - P_B)^2$ the usual Mandelstam variable. 

From the above equations, it is evident that measuring an azimuthally asymmetric distribution of pairs of unpolarized hadrons with modulation $\sin (\phi_{S_B}-\phi^{}_R)$ with respect to the collision plane, allows to isolate the term $A_{pp}$ where the chiral-odd IFF $H_1^{\sphericalangle\, c}$ is paired to the chiral-odd transversity $h_1^b$ and to the unpolarized parton distribution $f_1^a$. In fact, when a parton $a$ annihilates on the transversely polarized parton $b^\uparrow$, the polarization is transferred to the emerging parton $c^\uparrow$ which fragments into the observed hadron pair, while the other hadrons produced by the recoiling parton $d$ are summed over. Again, the analyzing power of the transverse polarization of the fragmenting parton is represented by the azimuthal orientation of the plane containing the final hadron pair momenta, and it is encoded in the IFF $H_1^{\sphericalangle\, c}$. All possible combinations of partons $a + b \rightarrow c + d$ must be included (see Ref.~\cite{Bacchetta:2004it} for the complete list), and they are described by the cross sections $d\hat{\sigma}$ and $d\Delta\hat{\sigma}$ for the unpolarized and polarized elementary $2\rightarrow 2$ processes, respectively. 


\begin{figure}[h]
\begin{center}
\resizebox{0.35\textwidth}{!}{%
  \includegraphics{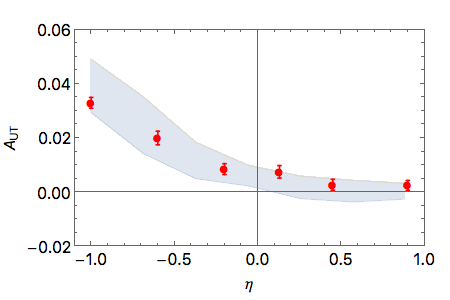}
}\\
\resizebox{0.5\textwidth}{!}{%
  \includegraphics{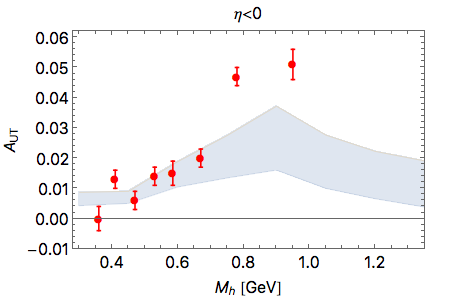} \includegraphics{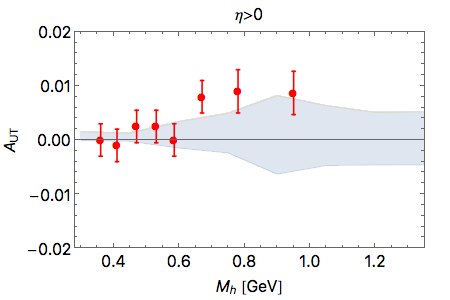}
}
\caption{The spin asymmetry $A_{pp}$ of Eq.~(\ref{eq::App}). Upper plot: $A_{pp}$ as function of $\eta$ integrated in $M_h$ and $|\bm{P}_{h\perp}|$. Lower plots: $A_{pp}$ as function of $M_h$ integrated in $|\bm{P}_{h\perp}|$ for $\eta < 0$ (left) and $\eta > 0$ (right). Data from the \texttt{STAR} measurement of Ref.~\cite{Adamczyk:2015hri} at $\sqrt{s} = 200$ GeV. Bands are the 68\% of 100 replicas for $h_1$ and $H_1^{\sphericalangle}$ from the analysis of Ref.~\cite{Radici:2015mwa}.}
\label{fig::App}
\end{center}
\end{figure}


The asymmetry $A_{pp}$ of Eq.~(\ref{eq::App}) has been measured by the \texttt{STAR} Collaboration for the process $p p^\uparrow \rightarrow (\pi^+ \pi^-) X$ at the cm energy of $\sqrt{s} = 200$ GeV~\cite{Adamczyk:2015hri}. Data are organized in a total of 16 bins covering $2\pi$ in azimuth for the central pseudorapidity region $-1 < \eta < 1$, the integrated luminosity is 1.8 pb$^{-1}$ with an average beam polarization of 60\%. The asymmetry $A_{pp}$ is extracted by fitting the $\sin (\phi_{S_B}-\phi^{}_R)$ modulation of the cross section; a very good $\chi^2$ / d.o.f. $\approx 1$ is reached. In Fig.~\ref{fig::App}, the $A_{pp}$ is considered after integrating over $|\bm{P}_{h\perp}|$. In the upper plot, it is shown as a function of $\eta$ after integrating also on $M_h$. In the lower plots, it is shown as a function of $M_h$ for $\eta < 0$ (left) and $\eta > 0$ (right). The negative pseudorapidities correspond to large $x$ in the valence region, where the transversity is larger. These data add a complementary and very useful information to what we already know on transversity from the SIDIS analysis, also for the higher statistical precision. An increase in the response is clearly visible for $M_h$ around the $\rho$ resonance mass. The bands represent a preliminary calculation of $A_{pp}$ using a 68\% of the 100 replicas for the transversity $h_1^q$ and the IFF $H_1^{\sphericalangle\, q}$ obtained in the SIDIS and $e^+ e^-$ analyses described in the previous sections, respectively. The preliminary nature of the calculations prevents from drawing any conclusion, but the agreement displayed in Fig.~\ref{fig::App} is definitely surprising and encouraging.

\section{Di-hadron observables at subleading twist}
\label{sec::twist3}

Higher-twist partonic functions describe multi-parton distributions corresponding to the interference of higher Fock components in the hadron wave function. Thus, they deliver  information on the physics of the largely unexplored quark-gluon correlations, which provide the energy that ultimately makes up the mass of the parent hadron. Moreover, higher-twist contributions are often necessary to correctly extract the leading-twist component from data obtained in the present kinematics of fixed-target experiments. 

The subleading-twist component (twist 3) of the di-hadron quark-quark correlator of Eq.~(\ref{eq::decomDelta}), when integrated over the quark $\bm{k}_T$ momentum, reads~\cite{Bacchetta:2003vn}
\begin{equation}
\begin{split}
\Delta_2 (z, &\zeta, M_h^2, \phi_R) = \frac{M_h}{16 \pi\, Q}\, \bigg\{ E + D^{\sphericalangle} \, \frac{\Rslash_T}{M_h} \\
&+ H \frac{i}{2}\, \Big[ \nslash_-, \, \nslash_+ \Big] + G^{\sphericalangle}\, \frac{\epsilon_T^{\rho \sigma} R^{}_{T \rho} \gamma^{}_{\sigma}}{M_h}\, \gamma_5 \bigg\} \; ,  
\end{split} 
\label{eq::decomDelta2}
\end{equation}
where $E$ and $H$ are chiral-odd functions, $H$ and $G^{\sphericalangle}$ are na\"ive time-reversal odd, and all DiFFs $E, \, D^{\sphericalangle}, \, H, \, G^{\sphericalangle}$ are functions of $(z,\, \zeta, \, M_h^2)$. 

\begin{figure}[h]
\begin{center}
\resizebox{0.25\textwidth}{!}{%
  \includegraphics{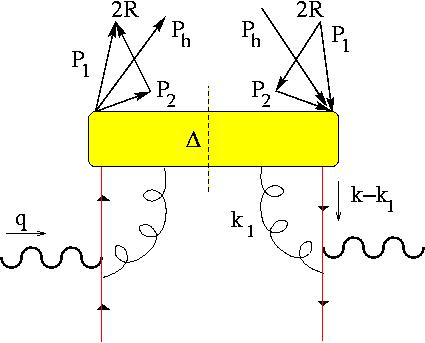}
}
\caption{Quark-gluon-quark correlation function $\Delta_A$ for the fragmentation of a quark with momentum $k$ into a pair of hadrons with momenta $P_1$ and $P_2$.}
\label{fig::DeltaA}       
\end{center}
\end{figure}

Consistently, the $\Delta_2$ correlator needs to be accompanied by the $1/Q-$suppressed quark-gluon-quark correlator depicted in Fig.~\ref{fig::DeltaA}, that is parametrized as~\cite{Bacchetta:2003vn}
\begin{equation}
\begin{split}
\Delta_A^\alpha (z, &\zeta, M_h^2, \phi_R) = \frac{M_h}{16 \pi\, z}\, \bigg\{ \tilde{D}^{\sphericalangle} \, \frac{R_T^\alpha}{M_h} \, \nslash_- - (\tilde{E} - i \tilde{H})\, \frac{\gamma^\alpha \, \nslash_-}{2} \\
&- i H_1^{\sphericalangle \, (1)} \, \frac{R_T^\alpha\, \Rslash_T}{M_h^2}\, \nslash_- + \tilde{G}^{\sphericalangle}\, \frac{\epsilon_T^{\alpha \beta} R^{}_{T \beta}}{M_h}\, \gamma_5 \, \nslash_- \bigg\} \; ,  
\end{split} 
\label{eq::decomDeltaA}
\end{equation}
where 
\begin{equation}
\begin{split}
\tilde{G}^{\sphericalangle} &= G^{\sphericalangle} - z\, G_1^{\perp\,(1)} - \frac{m}{M_h}\, z\, H_1^{\sphericalangle}\; , \quad \tilde{E} = E - \frac{m}{M_h}\,z\, D_1 \\
\tilde{D}^{\sphericalangle} &= D^{\sphericalangle} - z D_1^{(1)} \; , \qquad \tilde{H} = H + 2z \, H_1^{\perp\,(1)} \; , 
\end{split}
\label{eq:tildeDiFFs}
\end{equation}
and 
\begin{equation}
D_1^{(1)}(z,\zeta,M_h^2) = \int d\bm{k}_T\, \frac{\bm{k}_T^2}{2 M_h^2}\, D_1 (z, \zeta, M_h^2, \bm{k}_T^2, \bm{k}_T\cdot \bm{R}_T) \; ,
\label{eq::kTmom}
\end{equation}
and similarly for $G_1^{\perp\,(1)}, \, H_1^{\sphericalangle\, (1)}, \, H_1^{\perp\, (1)}$.  
The functions with tildes are all pure twist-3 objects, namely they disappear in the Wandzura-Wilzcek approximation. Again, they are all functions of $(z, \, \zeta, \, M_h^2)$. For each of them, there exists a partial-wave expansion similar to the one in Eq.~(\ref{eq::LMexp})~\cite{Bacchetta:2003vn}.

When computing the hadronic tensor for the SIDIS production of two unpolarized hadrons, the above $\Delta_2$ and $\Delta_A^\alpha$ are consistently combined with the corresponding correlators containing the PDFs at twist 2 and 3 levels. The resulting cross section contains several terms, depending also on the polarization state of the target and the lepton beam. For a longitudinally polarized lepton scattering off an unpolarized proton, the cross section contains an azimuthal $\sin\phi_R$ modulation~\cite{Bacchetta:2003vn} that can be isolated through the following beam-spin asymmetry:
\begin{align}
\lefteqn{A^{\text{LU}}_{\text{SIDIS}} (x, z, M_h; Q) =} \nn \\ 
&\quad  \frac{1}{\lambda}\, \frac{8}{\pi} \, \frac{\int d\phi_R \, d\cos\theta\,\sin \phi_R \, (d\sigma_{LU} - d\sigma_{-LU})}{\int d\phi_R \, d\cos\theta\, (d\sigma_{LU} + d\sigma_{-LU})} \nn \\
&= - \frac{W(y)}{A(y)} \,\frac{M}{Q}\, \frac{|\bm{R}|}{M_h} \nn \\
&\times 
\frac{\sum_q\,e_q^2\,\left[ x\, e^q(x)\, H_{1\, sp}^{\sphericalangle\, q}(z, M_h^2) + \frac{M_h}{z M}\, f_1^q (x)\, \tilde{G}_{sp}^{\sphericalangle\, q}(z, M_h^2) \right]} 
        {\sum_q\, e_q^2\, f_1^q(x) \, D_{1}^q(z, M_h^2)} 
\label{eq::SIDIS_long_bsa}
\end{align}
and longitudinal target-spin asymmetry 
\begin{align}
\lefteqn{A^{\text{UL}}_{\text{SIDIS}} (x, z, M_h; Q) =} \nn \\ 
&\quad  \frac{1}{|\bm{S}_L|}\, \frac{8}{\pi} \, \frac{\int d\phi_R \, d\cos\theta\,\sin \phi_R \, (d\sigma_{UL} - d\sigma_{U-L})}{\int d\phi_R \, d\cos\theta\, (d\sigma_{UL} + d\sigma_{U-L})} \nn \\
&= - \frac{V(y)}{A(y)} \,\frac{M}{Q}\, \frac{|\bm{R}|}{M_h} \nn \\
&\times 
\frac{\sum_q\,e_q^2\,\left[ x\, h_L^q(x)\, H_{1\, sp}^{\sphericalangle\, q}(z, M_h^2) + \frac{M_h}{z M}\, g_1^q (x)\, \tilde{G}_{sp}^{\sphericalangle\, q}(z, M_h^2) \right]} 
        {\sum_q\, e_q^2\, f_1^q(x) \, D_{1}^q(z, M_h^2)} \; .
\label{eq::SIDIS_long_tsa}
\end{align}
In the above equations, $V(y)=2 (2-y) \sqrt{1-y}$, $W(y) = 2y \sqrt{1-y}$, $\lambda$ is the lepton beam helicity which is flipped between positive $(d\sigma_{LU})$ and negative values $(d\sigma_{-LU})$, $|\bm{S}_L|$ is the similarly flipped longitudinal polarization of the target, and in each PDF and DiFF the dependence upon the hard scale $Q^2$ is understood. 

In Eqs.~(\ref{eq::SIDIS_long_bsa}) and~(\ref{eq::SIDIS_long_tsa}), the IFF $H_1^{\sphericalangle}$ appears as the chiral-odd partner of the chiral-odd PDFs $e(x)$ and $h_L(x)$. These PDFs are practically unknown from experiments. The former is of great importance because it is related to the soft physics of QCD chiral symmetry breaking. In fact, the isoscalar combination of first Mellin moments of $e(x)$ is related to the so-called pion-nucleon $\sigma-$term $\sigma_{\pi N}$. In QCD, the $\sigma-$term is related to matrix elements of the quark mass operator which explicitly breaks chiral symmetry. The $\sigma_{\pi N}$ is connected to the value at momentum transfer $t=0$ of the scalar form factor, which describes the elastic scattering off nucleon via the exchange of a spin-0 particle. The scalar form factor has not yet been measured except for its value in the time-like region at the so-called Chen-Dashen point $t=2m_\pi^2$, which can be deduced from pion-nucleon scattering data by means of low-energy theorems~\cite{80}. Chiral perturbation theory and dispersion relations allow to connect the experimental data in the time-like region to the $\sigma-$term in the space-like region~\cite{83}. The $\sigma_{\pi N}$ is also related to the strangeness content of the proton. Phenomenological values of $\sigma_{\pi N} = 50-70$ MeV~\cite{83} have sometimes been considered "large" because they would imply a large strange content in the nucleon. Lattice calculations of $\sigma_{\pi N}$ using the Feynman-Hellmann theorem bring results substantially compatible with phenomenology~\cite{95}. The second Mellin moment of $e^q(x)$ is proportional to the quark mass of a given flavor, thus offering access, in principle, to the current mass of quarks in DIS as well as to the scaling of this mass with dynamical chiral symmetry breaking. However, being $e^q(x)$ a subleading-twist PDF the effect is suppressed as $m_q / Q$. The third Mellin moment of $e(x)$ can be related to the average transverse force experienced by a transversely polarized quark in an unpolarized nucleon~\cite{Burkardtd2}. The first Mellin moment of $e(x)$ is the nucleon scalar charge.

On a wider perspective, a better understanding of the different nucleon charges can give hints into searches of new physics beyond the Standard Model. For example, elastic scattering of supersymmetric cold dark matter off nucleons depends on the $\sigma-$term $\sigma_{\pi N}$~\cite{DMsigmaterm}. More generally, model-independent bounds on direct dark matter detection include hadronic matrix elements of all bilinear operators, including scalar and tensor ones. Therefore, accumulating knowledge of their respective charges is of great importance.

The PDFs $e(x)$ and $h_L(x)$ appear in single-spin asymmetries also for single-hadron electroproduction. However, as for the case of transversity in the transverse target-spin asymmetry of Eq.~(\ref{eq::SIDISssa}), the DiFF formalism allows to work in collinear factorization also at subleading twist. Therefore, the expressions of the various spin asymmetries involve only products of PDFs and DiFFs, and not complicated convolutions on transverse momenta. In the specific case of Eqs.~(\ref{eq::SIDIS_long_bsa}) and~(\ref{eq::SIDIS_long_tsa}), the inclusive di-hadron SIDIS production gives a simpler direct access to $e(x)$ and $h_L(x)$, respectively, through the known chiral-odd leading-twist IFF $H_1^{\sphericalangle}$, without the need of introducing any additional model dependence on the quark transverse momentum $\bm{k}_T$. The main limitation to this strategy is represented in both cases by the term involving the DiFF $\tilde{G}^{\sphericalangle}$, which is totally unknown. The only available information is that in Wandzura-Wilzcek approximation this function vanishes. But this does not necessarily imply that its size should be negligible.

\subsection{The CLAS measurement}
\label{ssec::e_x_clas}

\begin{figure}[h]
\begin{center}
\resizebox{0.5\textwidth}{!}{%
  \includegraphics{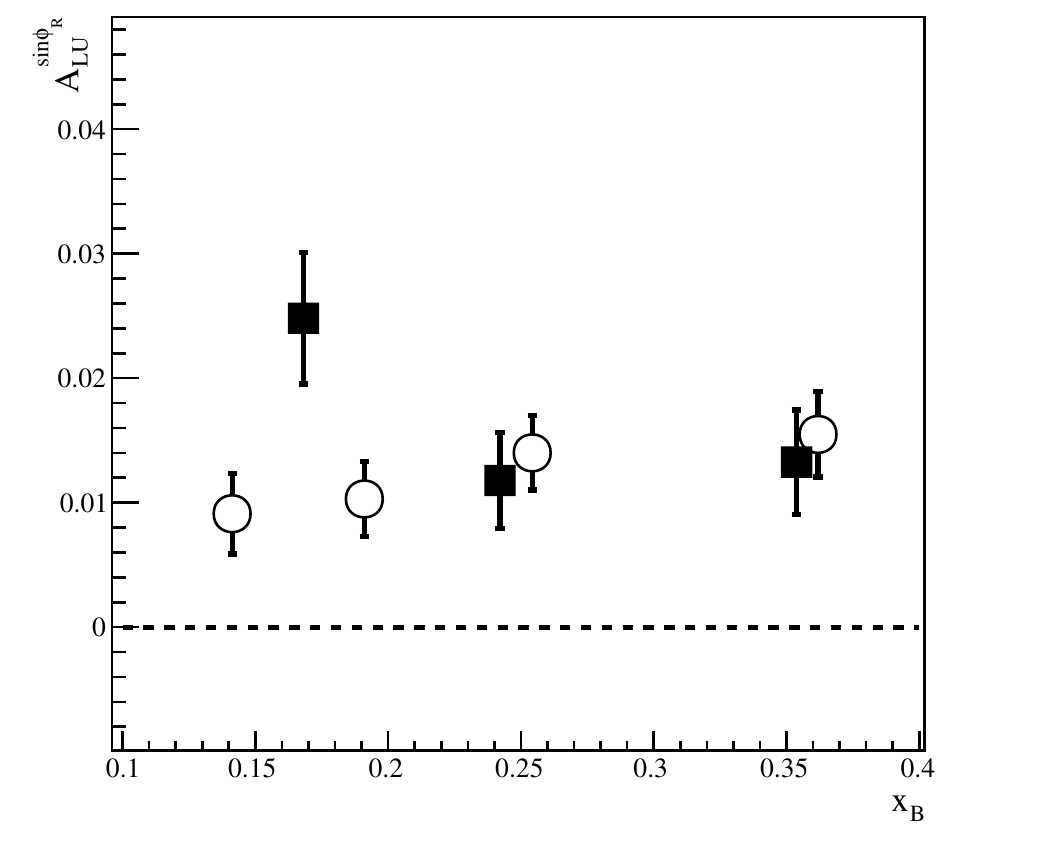} \includegraphics{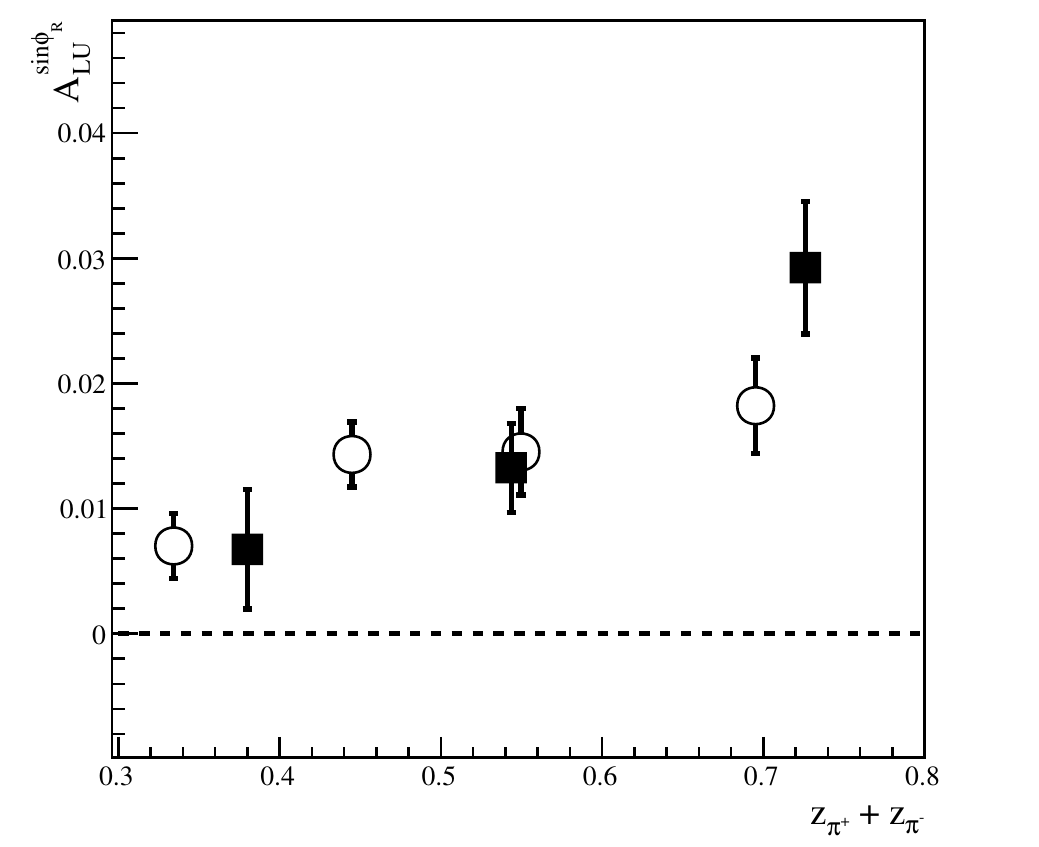} \includegraphics{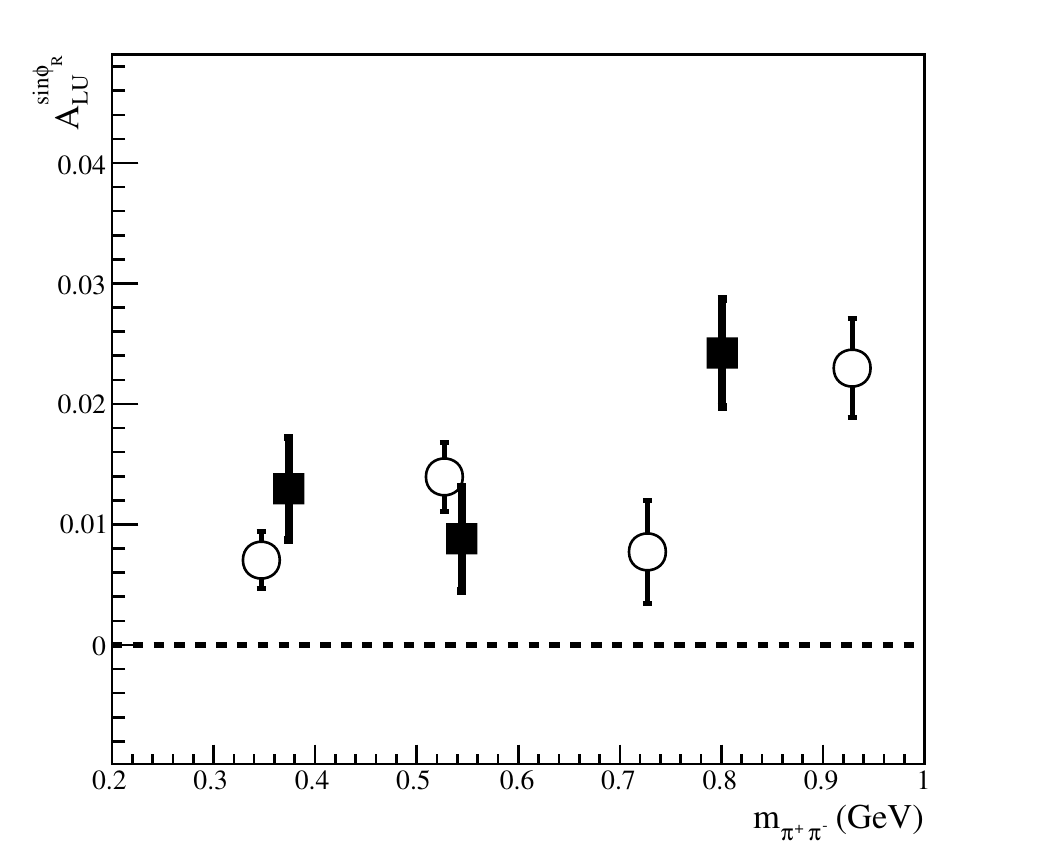}
}
\caption{The beam-spin asymmetry $A_{LU}$ of Eq.~(\ref{eq::SIDIS_long_bsa}) as a function of $x$ (left panel), $z$ (middle panel) and $M_h$ (right panel). Solid squares for an unpolarized hydrogen target~\cite{proc_menu2013_pisanos}, empty circles for a longitudinally polarized $NH_3$ target~\cite{proc_pereira_dis2015}.}
\label{fig::bsa_clas}
\end{center}
\end{figure}

A preliminary measurement of $A_{LU}$ for di-hadron SIDIS has been performed by the \texttt{CLAS} collaboration at Jefferson Lab for the process $e^{\to} p \rightarrow e' \pi^+ \pi^- X$~\cite{proc_menu2013_pisanos}. The extraction was performed on data collected by impinging a longitudinally polarized electron beam with an energy of 5.498 GeV on an unpolarized H$_2$ target. Events were selected through the cuts on the invariant mass $W^2 > 4$ GeV$^2$, the scale $Q^2 > 1$ GeV$^2$, the missing mass $M_X > 1.05$ GeV, and $y < 0.85$. The cut on the missing mass is used to remove the contribution from the exclusive production off a proton. Pions coming from the fragmentation of the target remnants are excluded by requesting positive values for the Feynman $x_F$ variable. An analogous measurement is carried on a longitudinally polarized NH$_3$ target, that aims at the extraction of both single and double-spin asymmetries~\cite{proc_pereira_dis2015}. In Fig.~\ref{fig::bsa_clas}, the extracted $A_{LU}$ is shown as a function of $x_B \approx x$ (left panel), $z$ (middle panel), and $M_h$ (right panel). The solid squares refer to results from the unpolarized hydrogen target, empty circles for the longitudinally polarized NH$_3$. The $A_{LU}$ is significantly different from zero in the whole kinematics explored. 
%
%

\begin{figure}[h]
\begin{center}
\resizebox{0.45\textwidth}{!}{%
  \includegraphics{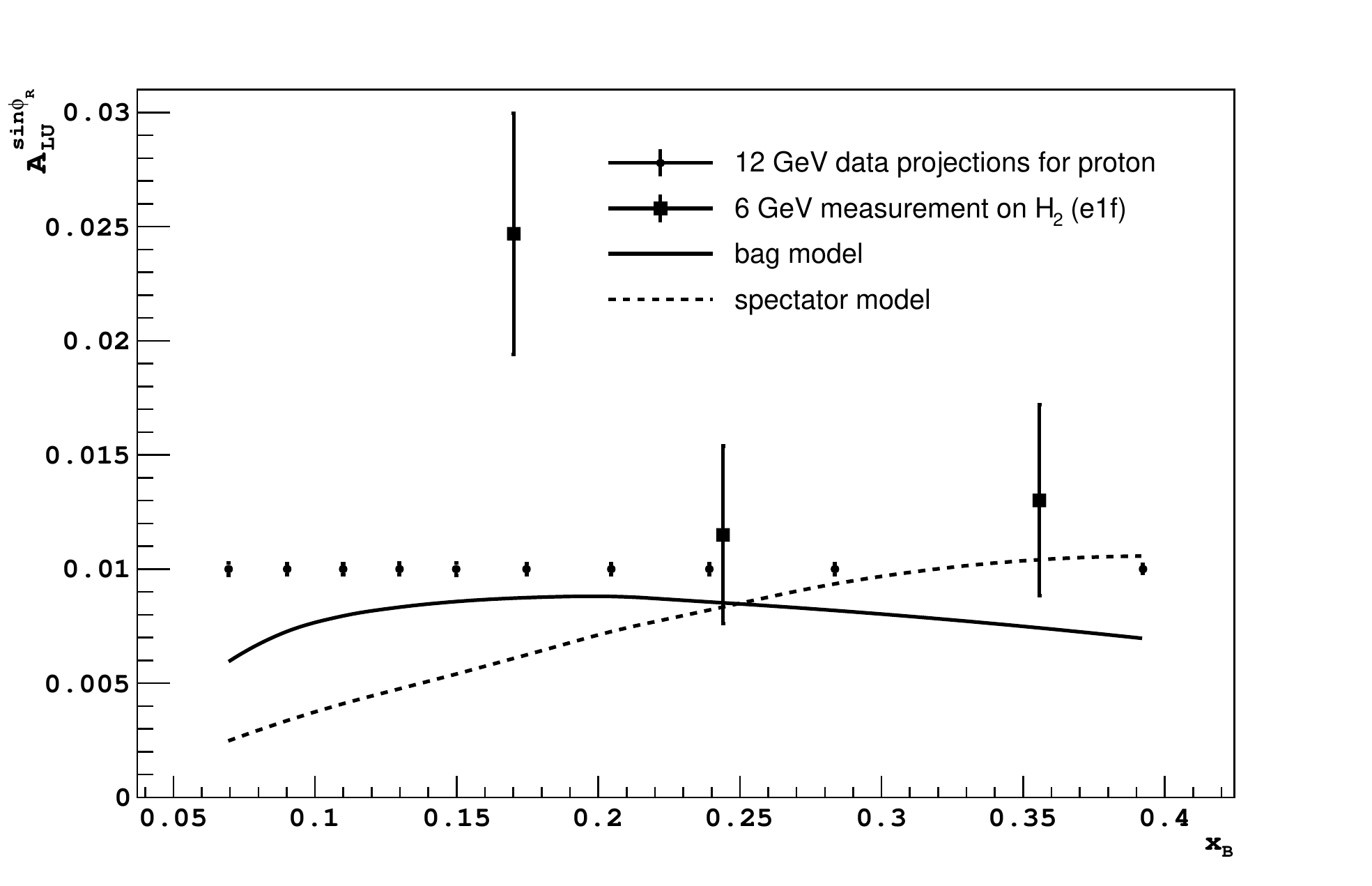}}\\
\resizebox{0.45\textwidth}{!}{%
  \includegraphics{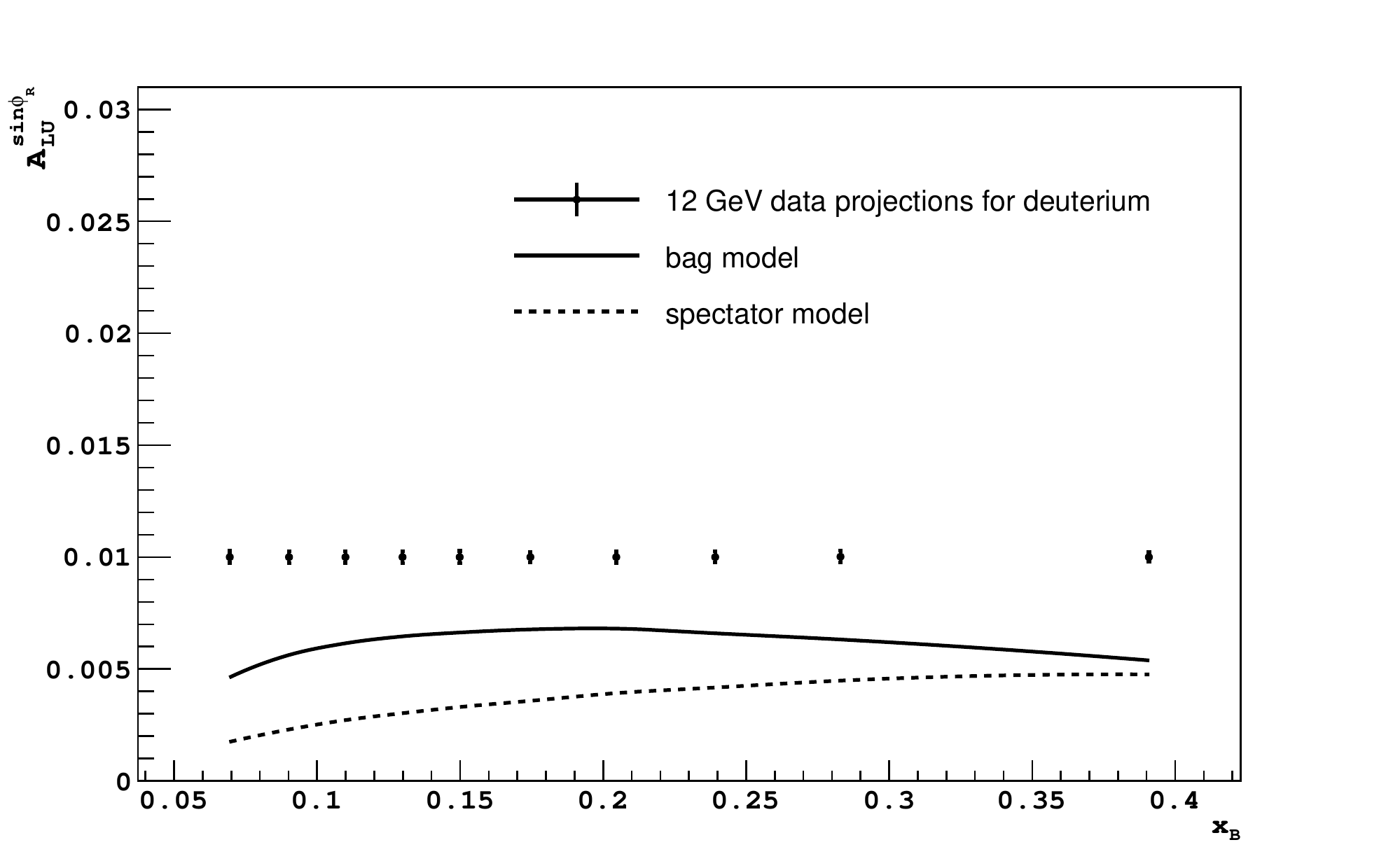}
}
\caption{Projections on errors for $A_{LU}$ as a function of $x_B \approx x$ with a 11-GeV lepton beam~\cite{prop_clas_11GeV}. Top panel: results for a proton target. Bottom panel: results for a deuterium target. In the top panel, the $A_{LU} (x)$ of Fig.~\ref{fig::bsa_clas} (solid squares) is also shown. Dashed and solid lines show the results when combining model predictions for $e(x)$ from the spectator~\cite{e_x_spectator_model} and bag~\cite{e_x_bag_model} models, respectively, with the IFF extracted in Ref.~\cite{Courtoy:2012ry}.}
\label{fig::bsa_clas_11GeV}
\end{center}
\end{figure}

An extension of this measurement has been recently approved at Jefferson Lab~\cite{prop_clas_11GeV} that will make use of the 11-GeV electron beam and of the new \texttt{CLAS12} detector. In Fig.~\ref{fig::bsa_clas_11GeV}, the top panel compares the results of previous measurement for $A_{LU}(x)$ (solid squares in the left panel of Fig.~\ref{fig::bsa_clas}) with the projected errors obtained with the upgrade to the a 11-GeV beam. The latter clearly will improve both the precision and the kinematic coverage. Moreover, it will extend the measurement also to a deuterium target, as shown in the bottom panel of Fig.~\ref{fig::bsa_clas_11GeV}. The two combined measurements will allow to separately extract $e^{u_v}(x)$ and $e^{d_v}(x)$. Because of the expected high statistics, data will be collected in all $(x_B, \, z, \, M_h)$ bins. The main advantage of such a 3-dimensional binning is that the $x$ dependence of the beam-spin asymmetry can be disentangled in a more accurate way, reflecting in a better knowledge of the $x-$dependence of the PDF $e^q$. It will also improve the analysis of a possible contribution from twist-3 DiFFs, whose dependence is expected to differ from the leading-twist one. 


\subsection{Extraction of $e(x)$}
\label{sec::e_x_extr}

A first attempt to extract $e(x)$ was performed in Ref.~\cite{efremov_e_x_ssa} using data for the SIDIS electro-production of a positive pion collected by the \texttt{CLAS} collaboration~\cite{bsa_piplus_clas}. As already stressed, in the context of single-hadron fragmentation the TMD factorization framework necessary to analyze the spin asymmetries implies model assumptions about the $\bm{k}_T$ dependence of the various TMD PDFs and FFs involved. Moreover, at the time of this analysis no parametrization was available for the Collins function $H_1^\perp(z)$, and the $z-$dependence of this chiral-odd partner of $e(x)$ had to be deconvoluted from the $z-$dependence of the transverse target-spin asymmetry by using a model calculation~\cite{efremov_z_aut}. 

\begin{figure}[h]
\begin{center}
\resizebox{0.4\textwidth}{!}{%
  \includegraphics{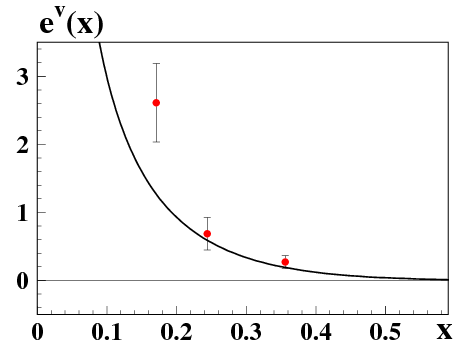}
}
\caption{The $e^V(x) = 4(e^{u_v}(x)-e^{d_v}(x))/9$ extracted from the preliminary \texttt{CLAS} data displayed in Fig.~\ref{fig::bsa_clas}~\cite{courtoy_e_x}. The solid line represents the LFCQM model prediction of Ref.~\cite{lfcqm_e_x}.}
\label{fig::ex_extraction}
\end{center}
\end{figure}

A new extraction has been recently performed in the context of collinear factorization using di-hadron fragmentation~\cite{courtoy_e_x}. It is based on the \texttt{CLAS} measurement of $A_{LU}$ on a hydrogen target~\cite{proc_menu2013_pisanos}, represented by the solid squares in Fig.~\ref{fig::bsa_clas}. Assuming the Wandzura-Wilzcek approximation and the symmetry properties of DiFFs, the $A_{LU}$ of Eq.~(\ref{eq::SIDIS_long_bsa}) is directly proportional to the flavor combination $e^V = (4 e^{u_v} - e^{d_v})/9$ through the IFF  $H_1^{\sphericalangle\, u}$, whose $z-$ and $M_h-$dependences were extracted in Ref.~\cite{Courtoy:2012ry} from the \texttt{BELLE} data~\cite{Vossen:2011fk}. Using the standard DGLAP evolution equations for DiFFs to scale $H_1^{\sphericalangle\, u}$ from \texttt{BELLE} to the energy of the \texttt{CLAS} measurement, the $z-$ and $M_h-$dependences of $A_{LU}$ can be integrated and its $x-$ dependence can be connected to the one of $e^V(x)$. The final result is shown in Fig.~\ref{fig::ex_extraction}. The solid line indicates the LFCQM model prediction of Ref.~\cite{lfcqm_e_x}, that appears in good agreement.


The biggest limitation to this extraction is the lack of information on the higher-twist fragmentation function $\tilde{G}^{\sphericalangle}$, that appears in the $A_{LU}$ of  Eq.~(\ref{eq::SIDIS_long_bsa}) and in the $A_{UL}$ of Eq.~(\ref{eq::SIDIS_long_tsa}) coupled to $f_1(x)$ and $g_1(x)$, respectively. The cross section for di-hadron production in $e^+e^-$ annihilations at subleading twist is not yet known. Thus, a possible strategies could be to study the ratio $A_{LU} / A_{UL}$. In fact, if the term proportional to $\tilde{G}^{\sphericalangle}$ would be negligible, using the symmetry properties of $H_1^{\sphericalangle\, q}$ the ratio should not exhibit any dependence on $(z, M_h)$, since the latter should cancel out between numerator and denominator. On the contrary, any observed dependence would hint at a non-negligible contribution from higher-twist fragmentation, making the extraction of $e(x)$ quite cumbersome. In this perspective, it is essential to collect high-precision data on the various observables in a common kinematics to be able to perform as accurate comparisons of their kinematical dependence as possible. 



\section{The TMD DiFFs}
\label{sec::TMDDiFF}

In the previous sections, we have explored the properties and the usefulness of DiFFs in the so-called collinear kinematics, {\it i.e.} when the dependence on the transverse momenta $\bm{k}_T$ of partons is integrated. In this limit, only two of the DiFFs listed in the leading-twist decomposition of Eq.~(\ref{eq::decomDelta}) survive, namely the unpolarized $D_1$ and the IFF $H_1^{\sphericalangle}$. We have examined their dependence on the pair fractional energy $z$ and pair invariant mass $M_h$ after performing a suitable expansion in relative partial waves of the hadron pair, and retaining only those components that survive the average on the left over $\cos\theta$ dependence. 

In this section, we briefly illustrate the potential of keeping the full dependence of DiFFs, namely of dealing with TMD DiFFs. In the next section, we consider in particular the helicity di-hadron fragmentation function $G_1^\perp (z, \zeta, \bm{R}_T^2, \bm{k}_T^2, \bm{k}_T \cdot \bm{R}_T)$.

\subsection{The helicity DiFF}
\label{sec::G1perp}

In Sec.~\ref{sec::e+e-ssa}, we have discussed the so-called Artru-Collins azimuthal asymmetry $A_{e^+e^-}$ that arises in the process $e^+ e^- \to (h_1,h_2)_{\text{jet1}}\, (\bar{h}_1,\bar{h}_2)_{\text{jet2}} X$ from the production of a correlated transversely polarized quark-antiquark pair. The azimuthal modulation $\cos(\phi_R +\bar{\phi}_R)\,\sin^2\theta_2$ (see also Fig.~\ref{fig::e+e-kin} for a definition of the angles) survives after integrating on the transverse total momenta of the hadron pairs, {\it i.e.} in the limit where each hadron pair is collinear with the direction of its related fragmenting quark and the pairs are emitted back-to-back. Collinear factorization framework allows to connect $A_{e^+e^-}$ to the simple product of two IFF $H_1^{\sphericalangle}$, one for each pair. 

In the collinear limit, no memory is kept of the transverse dynamics of fragmenting partons and both DiFFs $G_1^\perp$ and $H_1^\perp$ disappear after integrating on $\bm{k}_T$. In particular, the helicity DiFF $G_1^\perp$ vanishes because of parity invariance: if the fragmenting quark has momentum $\bm{k}$ and helicity $\bm{S}_{qL}$, and the hadron pair has a total momentum $\bm{P}_h$ collinear to $\bm{k}$, there is no further vector that allows to build a useful combination to represent the non-perturbative elementary mechanism in a way similar to the DiFF effect $\bm{S}_{qT}\cdot \bm{k} \times \bm{R}_T$ of Eq.~(\ref{eq::H1angle_dens}). Unless a suitable weighting function of $\bm{k}_T$ is introduced to preserve memory of the transverse parton dynamics. In fact, in Ref.~\cite{Boer:2003ya} it has been shown that in the same collinear limit the $e^+ e^-$ cross section contains also the azimuthal modulation $\cos 2(\phi_R -\bar{\phi}_R)$, whose coefficient gives the so-called longitudinal jet handedness azimuthal asymmetry
\begin{equation} 
\begin{split} 
&A_{e^+e^-}^{\sphericalangle}  =  \frac{1}{2 M_h^2 \bar{M}_h^2} \\
&\quad \times \, \frac{\sum_q e_q^2 \, G_{1 \sphericalangle}^{\perp\, q}(z, \zeta, \bm{R}_T^2; Q^2)\, \bar{G}_{1 \sphericalangle}^{\perp\, \bar{q}}(\bar{z}, \bar{\zeta}, \bar{\bm{R}}_T^2; Q^2)}
      {\sum_q e_q^2\, D_1^q (z, \zeta, \bm{R}_T^2; Q^2) \, \bar{D}_1^{\bar{q}} (\bar{z}, \bar{\zeta}, \bar{\bm{R}}_T^2; Q^2)} \; , 
\label{eq::e+e-jeta}
\end{split} 
\end{equation} 
where 
\begin{equation}
\begin{split} 
G_{1 \sphericalangle}^{\perp}(z, \zeta, \bm{R}_T^2; &Q^2) = \int d\bm{k}_T \, \bm{k}_T\cdot \bm{R}_T\\
&\times G_1^\perp (z, \zeta, \bm{R}_T^2, \bm{k}_T^2, \bm{k}_T\cdot \bm{R}_T; Q^2) \; .
\label{eq::momG1perp}
\end{split} 
\end{equation}
The $A_{e^+e^-}^{\sphericalangle}$ is the chiral-even counterpart of the Artru-Collins asymmetry. An analogous asymmetry involving chiral-even fragmentation functions does not emerge when only one hadron is detected in each jet. But this asymmetry can also be viewed as arising from the correlation of the longitudinal handedness functions of the two back-to-back jets. In fact, $(\bm{k}_T\times \bm{R}_T)\, G_1^\perp$ is proportional to the longitudinal jet handedness~\cite{Boer:2003ya} since it probes the helicity of the fragmenting quark. The asymmetry $A_{e^+e^-}^{\sphericalangle}$ cannot be directly translated to the handedness correlation observables defined in Ref.~\cite{Efremov:1992pe}. However, it may be interesting to study its behavior and search for possible deviations from standard expectations that could be due to $CP-$violating effects of the QCD vacuum~\cite{EfremovCPodd}. 

Similarly to Eq.~(\ref{eq::LMexp}), the helicity DiFF can be expanded in partial waves~\cite{Bacchetta:2002ux}
\begin{equation}
G_1^\perp \to G_{1, sp}^\perp + G_{1, pp}^\perp \, \cos\theta + G_{1, pp'}^\perp\, \sin\theta\, \hat{\bm{k}}_T\cdot \hat{\bm{R}} \; ,
\label{eq::LMG1perp}
\end{equation}
where each component is a function $G_{1, LL'}^\perp (z, \bm{R}_T^2, \bm{k}_T^2; Q^2)$.

By exploiting the $\zeta (\cos\theta)$ dependence in the cross section and by inserting the above expansion in Eq.~(\ref{eq::e+e-jeta}), the asymmetry $A_{e^+e^-}^{\sphericalangle}$ becomes
\begin{equation} 
\begin{split} 
&A_{e^+e^-}^{\sphericalangle}  =  2\pi^2\, |\bm{R}| \, |\bar{\bm{R}}|\, \sin^2\theta \, \sin^2\bar{\theta} \\
&\quad \times \, \frac{\sum_q e_q^2 \, G_{1, pp'}^{\perp\,(1)\, q}(z, M_h^2; Q^2)\, \bar{G}_{1, pp'}^{\perp\,(1)\, \bar{q}}(\bar{z}, \bar{M}_h^2; Q^2)}
      {\sum_q e_q^2\, D_1^q (z, M_h^2; Q^2) \, \bar{D}_1^{\bar{q}} (\bar{z}, \bar{M}_h^2; Q^2)} \; , 
\label{eq::e+e-jetaLM}
\end{split} 
\end{equation} 
where
\begin{equation}
G_{1, pp'}^{\perp\,(1)}(z, M_h^2; Q^2) = \int d\bm{k}_T \, \frac{\bm{k}_T^2}{2 M_h^2}\, G_{1, pp'}^\perp (z, \bm{R}_T^2, \bm{k}_T^2; Q^2) \; .
\label{eq::momLMG1perp}
\end{equation}

By summing Eq.~(\ref{eq::e+e-jetaLM}) over one emisphere, the \texttt{BELLE} collaboration has investigated the $z-$ and $M_h-$dependence of $A_{e^+e^-}^{\sphericalangle}$ by extracting the $\cos 2(\phi_R -\bar{\phi}_R)\, \sin^2\theta$ modulation of the cross section for the case of the production of two back-to-back $(\pi^+ \pi^-)$ pairs. Unexpectedly, in both cases the asymmetry is zero within the experimental error~\cite{BELLEG1perp}. Since there is no apparent compelling reason for the $G_{1, pp'}^\perp$ function to vanish, this surprising experimental evidence needs further investigations. 

\subsection{The SIDIS cross section}
\label{sec::2hSIDISTMD}

The cross section of Eq.~(\ref{eq::SIDIScross}) describes the particular case of di-hadron SIDIS production in single-photon-exchange approximation and at leading twist, when the kinematics is collinear, the lepton beam is unpolarized, and the target is transversely polarized. For generic polarization states $X$ and $Y$ of the beam and target, respectively, the cross section is differential in $dx$, $dy$, $dz$, $d\bm{P}_{h\perp}^2$, $d\phi_h$, $dM_h^2$, $d\phi_R$, $d\cos\theta$, $d\phi_S$, where the transverse components and the azimuthal angles are measured in the plane where $(P, q)$ are collinear (see Fig.~\ref{fig::SIDISkin}). Its general expression is given by~\cite{Gliske:2014wba}
\begin{equation}
\begin{split}
&d\sigma_{XY} = K_{XY} (x,y; Q^2)\\
&\quad \times \sum_{\ell = 0}^{\ell_{\text{max}}} \sum_{m = -\ell}^\ell\, P_{\ell m} (\cos\theta) \, f_m (\cos\phi_h, \, \cos\phi_R)\, F_{XY}^{P_{\ell m}\, f_m} \; , 
\label{eq::SIDIS_XY}
\end{split}
\end{equation}
where $K_{XY}$ is a phase space factor, $P_{\ell m}$ are Legendre polynomials, $f_m$ are trigonometric functions of the azimuthal angles, and $F_{XY}^{P_{\ell m}\, f_m}$ are structure functions of $x$, $z$, $M_h^2$, $\bm{P}_{h\perp}^2$, and $Q^2$. The sum runs over $\ell = L \oplus L'$, where $L, \, L'$ are the relative partial waves for each hadron pair. 

Since there are two sources of angular momentum (the total momentum $P_h$ and the relative momentum $R$ of the hadron pair) and the hadronic tensor is not necessarily linear in $R$, the sum on partial waves is unlimited. The only constrain is that the sum of the coefficients of $\phi_h$ and $\phi_R$ in the various $f_m$ functions is bounded to at most 3, because this is the maximum mismatch of angular momentum projections in the virtual-photon-proton system. If the hadron pair invariant mass is limited at $M_h \lesssim 1$ GeV, then $\ell_{\text{max}} = 2$. 

If this range of invariant masses satisfies the condition $M_h \ll Q$, the TMD factorization theorems for single-hadron fragmentation at leading twist~\cite{Ji:2004wu,Collins:2004nx,Echevarria:2012pw,Collins:2012uy} can be extended to the di-hadron fragmentation case. Assuming the same framework also at subleading twist~\cite{Bacchetta:2006tn}, we can parametrize the structure functions $F_{XY}^{P_{\ell m}\, f_m}$ as convolutions of TMD PDFs and TMD DiFFs:
\begin{equation}
\begin{split}
&F_{XY}^{P_{\ell m}\, f_m} (x, z, M_h^2, \bm{P}_{h\perp}^2) = \\
&\quad \sum_q e_q^2 \int d\bm{k}_\perp \, d\bm{P}_{hT}\, \delta (z \bm{k}_\perp + \bm{P}_{hT} - \bm{P}_{h\perp}) \\
&\qquad \times w_{XY} (x, z, M_h, \bm{k}_\perp, \bm{P}_{hT})\  \text{PDF} (x, \bm{k}_\perp) \\
&\qquad \times \text{DiFF}_{\ell m} (z, M_h^2, |\bm{P}_{hT}|) \; , 
\label{eq::FXY}
\end{split}
\end{equation}
with $\bm{k}_\perp$ the parton transverse momentum in the $(P, q)$ collinear plane, $\bm{P}_{hT}$ the total transverse momentum of the hadron pair with respect to the fragmenting quark direction, and $w_{XY}$ is a suitable weighting function. With this formalism, the transverse-momentum-dependent cross section for di-hadron SIDIS production up to subleading twist and for any polarization state of lepton beam and target, has been presented for the first time in Ref.~\cite{Gliske:2014wba}. It contains and recovers all the leading-twist contributions discussed in the previous literature (see, e.g., Ref.~\cite{Bacchetta:2002ux}). 

Among the various terms, here we mention the leading-twist structure function $F_{UL}^{P_{\ell m} \sin m(\phi_h-\phi_R)}$ that contains the convolution of the helicity distribution $g_1$ with the helicity DiFF $G_1^\perp$ discussed in the previous section. Since $g_1$ is known to considerable accuracy, one can extract $G_1^\perp$ from the $e p^\to \to (h_1, h_2) X$ cross section and actually predict the longitudinal jet handedness correlation in $e^+ e^- \to (h_1, h_2) (\bar{h}_1, \bar{h}_2) X$: any experimental deviation could be interpreted as a $CP-$violating effect of the QCD vacuum~\cite{EfremovCPodd}. Moreover, the helicity DiFF occurs also in the leading-twist $F_{UT}^{P_{\ell m} \sin ((1+m)\phi_h + m\phi_R - \phi_S)}$ convoluted with the TMD PDF $g_{1T}$. This function (extrapolated at $x=0$) gives information on violations of the Burkhardt-Cottingham sum rule. 

Finally, we mention that in Ref.~\cite{Kotzinian:2014lsa} a definition of $\phi_R$ different from Eq.~(\ref{eq::phiR}) is adopted. It leads to the transverse component of $R$ in the target rest frame equal to $\bm{R}_\perp = (\bm{P}_{1\perp} - \bm{P}_{2\perp})/2$. The covariant definition of $\phi_R$ in Ref.~\cite{Gliske:2014wba} instead brings to $\bm{R}_{T\perp} = (z_2 \bm{P}_{1T} - z_1 \bm{P}_{2T})/z$, which coincides with the expressions adopted in the various experimental extractions of transversity up to $1/Q^2$ corrections~\cite{Gliske:2014wba}. The main difference between the two definitions emerges when the hadron pair is collinear with the fragmenting quark, {\it i.e.} for $\bm{P}_{hT} = 0$: with the latter, $\bm{R}_T$ is disconnected from the quark transverse momentum $\bm{k}_T$; with the former, there is a direct relation~\cite{Kotzinian:2014gza}. Hence, with this definition the $\bm{P}_{hT}-$integrated cross section still displays a $\phi_R-$modulation related to the Sivers effect: in a transversely polarized nucleon, the azimuthal orientation of the hadron pair plane in momentum space is sensitive to spin-orbit correlations inside the nucleon. This sensitivity has been widely explored with Monte Carlo simulations of the Sivers effect in di-hadron production at various kinematical setups of interest~\cite{Matevosyan:2015gwa}, using the NJL-jet model at each vertex of the quark hadronization chain~\cite{Casey:2012ux}.

\section{Summary and outlooks}
\label{sec::conclusions}

Di-hadron fragmentation functions (DiFFs) describe the direct fragmentation of a (polarized) parton into a pair of hadrons. They can be extracted from data for electron-positron annihilations where two back-to-back jets are produced and a pair of hadrons is detected in each jet. When the pair is collinear with the jet axis (or, equivalently, with the fragmenting quark momentum), only the two DiFFs $D_1$ and $H_1^\sphericalangle$ survive, describing the fragmentation of an unpolarized or transversely polarized parton, respectively. Using the \texttt{BELLE} measurement of the Artru-Collins azimuthal asymmetry~\cite{Vossen:2011fk}, the dependence of $H_1^\sphericalangle$ on the pair's fractional energy $z$ and invariant mass $M_h$ was parametrized for the first time from data~\cite{Courtoy:2012ry}. The chiral-odd $H_1^\sphericalangle$ describes a new non-perturbative mechanism where the azimuthal orientation of the hadron pair in momentum space can play the role of spin analyzer of the transverse polarization of the fragmenting quark. Besides this, the extraction of this chiral-odd DiFF opened the way to a more convenient access to the transversity distribution $h_1$, the missing piece in a complete picture of the collinear spin structure of the nucleon at leading twist. In fact, the simple product $h_1 \ H_1^\sphericalangle$ can be isolated in the leading-twist cross section for semi-inclusive electro-production of two hadrons through a transverse target-spin asymmetry, with no need to specify any dependence on the transverse momentum of partons. 

The first collinear extraction of transversity was realized in Ref.~\cite{Bacchetta:2011ip} by combining the \texttt{BELLE} data with the electro-production data on a proton target from the \texttt{HERMES} collaboration~\cite{hermes_2008}. Later, using also the \texttt{COMPASS} data on proton and deuteron targets~\cite{compass_2012,compass_2014}, the valence $u$ and $d$ components of transversity could be separated~\cite{Bacchetta:2012ty,Radici:2015mwa}, resulting in reasonable agreement with the extraction based on the Collins effect in single-hadrons fragmentation~\cite{Anselmino:2013vqa,Kang:2015msa}. The same combination $h_1 \ H_1^\sphericalangle$ happens also in the leading-twist cross section for di-hadron production in hadronic collisions when one of the two hadrons is transversely polarized~\cite{Bacchetta:2004it}. Recently released data for the related spin asymmetry by the \texttt{STAR} collaboration~\cite{Adamczyk:2015hri} are in very good agreement with the preliminary predictions based on transversity and DiFFs extracted from elsewhere, suggesting that these partonic functions are indeed universal. 

The knowledge of transversity is limited to a restricted range in the fractional parton momentum $x$, thus preventing from a reliable calculation of its first Mellin moment, the nucleon tensor charge, whose knowledge would help in the exploration of new observables sensitive to interactions with dark matter~\cite{DMsigmaterm,DelNobile:2013sia} or to new physics beyond the Standard Model~\cite{Cirigliano:2013xha}. An extension of the existing electro-production measurements is planned at Jefferson Lab during the realization of the 12-GeV program for both proton~\cite{dh_clas12_p} and effective neutron~\cite{dh_solid_n} targets. This will provide high-precision data for separate $u$ and $d$ flavors at larger $x$ in the valence region. Using the same \texttt{CLAS12} detector, another proposal has been approved~\cite{prop_clas_11GeV} to measure $(\pi^+ \pi^-)$ multiplicities and directly extract the unpolarized DiFF $D_1$ which, at the moment, is parametrized from the output of the \texttt{PYTHIA} Monte Carlo adapted to the \texttt{BELLE} kinematics. The \texttt{COMPASS} collaboration is also analyzing the same observable and some preliminary results have been reported in Ref.~\cite{Makke:2013dya}. 

The DiFFs play also a role in extending the knowledge of the nucleon collinear picture beyond the leading twist. The same chiral-odd $H_1^\sphericalangle$ provides the cleanest access to the poorly known twist-3 parton distributions $e(x)$ and $h_L(x)$~\cite{Bacchetta:2003vn}, which are directly connected to quark-gluon correlations. In particular, the $e(x)$ is intimately related to the mechanism of dynamical chiral symmetry breaking in QCD through the isoscalar combination of its Mellin moments, which is proportional to the pion-nucleon $\sigma$-term. A preliminary measurement of the related di-hadron beam-spin asymmetry has been performed by the \texttt{CLAS} collaboration~\cite{proc_menu2013_pisanos}, leading to a preliminary extraction of $e(x)$~\cite{courtoy_e_x} in good agreement with model calculations. This measurement will be improved, both in precision and kinematical coverage, during the upcoming 12-GeV program at Jefferson Lab~\cite{prop_clas_11GeV}. 

The DiFFs can be a useful tool also when keeping information on the transverse momentum dynamics of partons; in this case, we speak of TMD DiFFs. For example, the cross section for the electro-production of two hadrons has a very rich structure~\cite{Gliske:2014wba} and it is easy to come across terms that are similar to the single-hadron fragmentation case, and whose measurement can represent an important cross-check of the elementary mechanism described by the corresponding TMD PDF. Moreover, we can find contributions that have no such counterpart. The chiral-even helicity TMD DiFF $G_1^\perp$ is responsible for the so-called longitudinal jet handedness azimuthal asymmetry in $e^+ e^-$ annihilations, that has no analogous one in single-hadron fragmentation~\cite{Boer:2003ya}. The $G_1^\perp$ can be connected to the longitudinal jet handedness to explore possible effects due to $CP-$violation of the QCD vacuum~\cite{EfremovCPodd}.

Despite this large set of results and measurements, the nucleon partonic structure still remains largely unexplored, particularly at small values of parton fractional momenta where non-valence degrees of freedom are predominant. New explorations are needed in this kinematical domain that hopefully will become possible with the advent of an Electron-Ion Collider machine (\texttt{EIC}). In that context, di-hadron fragmentation functions will certainly continue to play a major role in the investigations.

%

\end{document}